\documentclass[fleqn,usenatbib]{mnras}

\usepackage{newtxtext,newtxmath}

\usepackage[T1]{fontenc}

\DeclareRobustCommand{\VAN}[3]{#2}
\let\VANthebibliography\thebibliography
\def\thebibliography{\DeclareRobustCommand{\VAN}[3]{##3}\VANthebibliography}

\usepackage{graphicx}	
\usepackage{changepage}
\usepackage{amsmath}	

\title[Simulating Galaxy Surveys with CASTOR]{FORECASTOR -- II. Simulating Galaxy Surveys with the Cosmological Advanced Survey Telescope for Optical and UV Research}

\author[M. A. Marshall et al.]{Madeline A. Marshall,$^{1,2}$\thanks{Email: \href{mailto:mmarshall@lanl.gov}{mmarshall@lanl.gov}} Laurie Amen,$^{1,3}$ Tyrone E.~Woods,$^{1,4}$ Patrick C{\^o}t{\'e},$^1$ L.~Y.~Aaron {Yung},$^{5,6}$ 
\newauthor
Melissa Amenouche,$^1$ Emily K. Pass,$^{7,8}$ Michael L. Balogh,$^{8,9}$ Samir Salim,$^{10}$ and Thibaud Moutard$^{11,12}$\\
$^1$National Research Council of Canada, Herzberg Astronomy \& Astrophysics
Research Centre,
5071 West Saanich Road, Victoria, BC V9E 2E7, Canada\\
$^2$Los Alamos National Laboratory, Los Alamos, NM 87545, USA\\
$^3$Department of Physics and Trottier Space Institute, McGill University, 3600
University Street, Montr{\'e}al, QC H3A 2T8, Canada\\
$^4$Department of Physics and Astronomy, Allen Building, 30A Sifton Rd,
University of Manitoba, Winnipeg, MB  R3T 2N2, Canada\\
$^5$Astrophysics Science Division, NASA Goddard Space Flight Center, 8800 Greenbelt Rd, Greenbelt, MD 20771, USA\\
$^6$Space Telescope Science Institute, 3700 San Martin Drive, Baltimore, MD 21218, USA\\
$^7$Center for Astrophysics $\vert$ Harvard \& Smithsonian, 60 Garden Street, Cambridge, MA 02138, USA\\
$^8$Department of Physics and Astronomy, University of Waterloo, Waterloo, Ontario, N2L 3G1 Canada\\
$^9$Waterloo Centre for Astrophysics, Waterloo, Ontario, N2L 3G1 Canada\\
$^{10}$Department of Astronomy, Indiana University, Bloomington, IN 47405, USA\\
$^{11}$European Space Agency (ESA), European Space Astronomy Centre (ESAC), Camino Bajo del Castillo s/n, 28692 Villanueva de la Cañada, Madrid, Spain\\
$^{12}$Aix Marseille Univ, CNRS, CNES, LAM, Marseille, France\\
}

\date{Accepted 2024 December 20. Received 2024 November 26; in original form 2024 July 16}

\pubyear{\the\year{}}

\begin{document}
\label{firstpage}
\pagerange{\pageref{firstpage}--\pageref{lastpage}}
\maketitle

\begin{abstract}
The Cosmological Advanced Survey Telescope for Optical and UV Research (CASTOR) is a planned flagship space telescope, covering the blue-optical and UV part of the spectrum. Here we introduce the CASTOR image simulator, a Python GalSim package-based script capable of generating mock CASTOR images from an input catalogue. We generate example images from the CASTOR Wide, Deep, and Ultra-Deep surveys using simulated light-cones from the Santa Cruz Semi-Analytic Model. We make predictions for the performance of these surveys by comparing galaxies that are extracted from each image using Source Extractor to the input catalogue. 
We find that the Wide, Deep, and Ultra-Deep surveys will be 75\% complete for point sources down to $\sim27$, 29 and 30 mag, respectively, in the UV, u, and g filters, with the UV-split and u-split filters reaching a shallower depth. With a large area of $\sim2200$ deg$^2$, the Wide survey will detect hundreds of millions of galaxies out to $z\sim4$, mostly with $M_\ast\gtrsim10^9M_\odot$. The Ultra-Deep survey will probe to $z\sim5$, detecting galaxies with $M_\ast\gtrsim10^7M_\odot$.
These galaxy samples will enable precision measurements of the distribution of star formation in the cosmic web, connecting the growth of stellar mass to the assembly of dark matter halos over two thirds of the history of the Universe, and other core goals of CASTOR's legacy surveys.
These image simulations and the tools developed to generate them will be a vital planning tool to estimate CASTOR’s performance and iterate the telescope and survey designs prior to launch.

\end{abstract}



\section{Introduction} \label{sec:intro}

Space astronomy has entered an extraordinary era of deep, high-resolution capabilities across the optical--infrared, with the high resolution of the James Webb Space Telescope \citep[JWST;][]{Gardner2006,Gardner2023,McElwain2023} complemented by new wide field optical--infrared observatories, including ESA's new Euclid mission \citep{Laureijs2011,Racca2016} and NASA's upcoming Roman Space Telescope \citep{Spergel2013, Spergel2015}.
However, long term astronomical planning exercises around the world \citep[e.g., Astro2020, LRP2020, Voyage 2050;][]{Astro2020, LRP2020, Voyage2050} have identified a looming gap in the astronomical community's space-based ultraviolet and blue-optical capability. To avoid astronomy's capacity to observe in the UV lagging far behind other wavebands, there is an urgent need for a high sensitivity, high angular resolution, wide field of view instrument in the next decade.

The Cosmological Advanced Survey Telescope for Optical and UV Research (CASTOR) is a planned 1m-class UV/optical telescope led jointly by the National Research Council of Canada and the Canadian Space Agency, in bilateral partnership with the United Kingdom Space Agency.
CASTOR will offer simultaneous UV, u, and g-band imaging with $\sim 0\farcs15$ angular resolution over a 0.25 deg$^2$ field of view, with a filter wheel including a bandpass filter and a grism, as well as UV multi-object spectroscopy in a parallel field \citep{CASTOR_SMS, CASTOR}. Enabled by these unique capabilities, CASTOR's planned surveys will explore the evolution of galaxies at Cosmic Noon, probe the origins of cosmic explosions, unlock a new era of precision cosmology, test the habitability of distant worlds, and unveil the nature of everything from the composition of icy bodies in the outer solar system to the masses of distant quasars.

After over a decade of development, the CASTOR mission concept has recently completed its Phase 0 science planning, including the initial design of its primary and legacy surveys \citep{Phase0Science}. A core component of this work has been the construction of detailed mission planning and science simulation tools for individual instruments \citep[e.g.,][]{UVMOS, Cheng2023}, as well as each of the science surveys, collectively the Finding Optical Requirements and Exposure times for CASTOR (FORECASTOR) project. 
With FORECASTOR we will supply a suite of planning tools for observation preparation, data reduction, and analysis for the CASTOR mission. 
In \citet{Cheng2023} we presented the first of these tools, the photometric exposure time calculator (ETC). 
With the ETC, users can find the estimated exposure time required to achieve a given signal-to-noise rato (S/N), or the converse, for a specified input source, for each of CASTOR's photometric filters: UV, UV-split, u, u-split, and g. This public tool will be instrumental in the planning of specific science observations. 

In this work we aim to extend these photometric predictions beyond individual sources by characterizing the planned extragalactic surveys on a statistical level.
We aim to estimate the capabilities of the CASTOR telescope for carrying out CASTOR's planned Wide, Deep, and Ultra-deep Legacy Surveys by making mock images of extragalactic fields. By testing galaxy extraction on these images, we estimate the completeness of the various CASTOR surveys, relative to a galaxy's magnitude, stellar mass, radius, and redshift. This gives a more detailed understanding of the capabilities of this instrument. These mock images can also be used in future analyses for determining the performance of CASTOR under specific science use cases.

This paper is outlined as follows. In Section \ref{sec:Surveys} we describe the Wide, Deep and Ultra-Deep CASTOR Legacy Surveys.
In Section \ref{sec:Methods} we describe our methodology of creating the mock images, and the input galaxy catalogues that we use within the images.
In Section \ref{sec:Images} we show examples of our mock CASTOR images, and how they compare with other instruments.
In Section \ref{sec:Results} we investigate the completeness of the various CASTOR surveys as measured from these mock images.
We include a discussion in Section \ref{sec:Discussion}, and finally we conclude in Section \ref{sec:Conclusions}.
We assume a Lambda-CDM cosmology with $H_0=67.8$ km / (Mpc s), $\Omega_m=0.308$, and $\Omega_\Lambda=0.692$ \citep{Planck2015}.
 
\section{Planned CASTOR Extragalactic Surveys}
\label{sec:Surveys}

Within the context of galaxy surveys, three key legacy surveys planned for CASTOR are the Wide, Deep and Ultra-Deep surveys, which together comprised a tiered ``wedding-cake'' approach to CASTOR's galaxy evolution science (a fourth survey, the CASTOR Nearby Galaxies Survey, is beyond the scope of this work). 
The exposure times and survey areas of these surveys are summarized in Table \ref{tab:Surveys}.
The Wide Survey will image the 2227 deg$^2$ Roman High Latitude Wide Area Survey (HLS) field to a point source depth of $\sim$27--28 mag; the Deep Survey will image 83 deg$^2$ in six contiguous regions overlapping the Rubin/Euclid Deep fields, reaching a point source depth of $\sim29$ mag; and finally, the Ultra-Deep survey will span four CASTOR pointings, totalling 1 deg$^2$, to a depth of $\sim30$--31 mag (in this paper we predict completeness limits as given in Table \ref{tab:CompletenessLims}). Together, these surveys will trace the distribution of star formation within the cosmic web, enable precision photometric and spectroscopic redshift measurements, connect the growth of stellar mass over cosmic time to the assembly of dark matter halos, probe the outer reaches of massive, quiescent galaxies, identify a large sample of faint, post-starburst galaxies, and extend star formation rate and morphology measurements of dwarf galaxies out to high redshifts.

Each of the surveys will be conducted across the five photometric bandpasses, UV, UV-split, u, u-split, and g.
The 3-mirror anastigmat optical design of CASTOR results in the light being split between the three filter channels (UV, u, and g). An additional filter can be placed before the UV and u band detectors, resulting in the u-split and UV-split bandpasses. When observing with and without the splitting filter, to first get the UV and u images, and then the UV-split and u-split images, the g-band is observed twice, doubling its exposure time.
Thus, in each of the surveys the g-band will be observed with twice the exposure time as the other filters. 
For more details about the specific filters, see \citet{Cheng2023}.

The Wide survey will have a total exposure time of 1000s in the UV, UV-split, u, and u-split filters, while the g-band will have an exposure time of 2000s. This will enable a vast range of cosmological studies, breaking degeneracies in measuring photometric redshifts, deblending sources and carrying out shape measurements, and studying tens of thousands of strong lenses, while yielding a vast treasury of galaxies within a footprint overlapping the Rubin Observatory’s Legacy Survey of Space and Time (LSST) and Euclid's surveys. 
The Deep survey will have a total exposure time of 18,000s in each filter, except 36,000s for the g-band. It will detect millions of galaxies at $z<1.5$ and, combined with the great wealth of ancillary data available from the optical to radio, will allow us to connect the in-situ growth mechanisms of galaxies to their location in the cosmic web.
The Ultra-Deep survey will have ten times the exposure time of the Deep survey, with a total exposure time of 180,000s per filter, with 360,000s for the g-band. It will extend the Deep survey's science to $z>1.5$, and allow for pixel-by-pixel star formation rate estimates through spectral energy distribution (SED) fitting of distant galaxies.
The goal of this work is to provide detailed estimates of the performance of each of these surveys.

\begin{table}
\caption{The exposure times and survey areas for the planned CASTOR Wide, Deep and Ultra-Deep surveys. The g-band exposures have twice the exposure time as the UV, UV-split, u, and u-split filters.
}
\label{tab:Surveys}
\begin{tabular}{cccc}
\hline 
\hline 
Survey & \multicolumn{2}{c}{Exp. Time (s)} & Area (deg$^2$) \\
& \footnotesize{UV, UV-split, u, u-split} & g\\\hline
Wide & 1000 & 2000 & 2227 \\
Deep & 18000 & 36000 & 83 \\
Ultra-Deep & 180000 & 360000& 1 \\
\hline 
\end{tabular} 
\end{table}

\section{Simulating Deep Fields}
\label{sec:Methods}

\subsection{GalSim}
\label{sec:GalSim}

GalSim \citep{2015A&C....10..121R} is an open-source Python package that allows the user to simulate images of astrophysical objects (stars and galaxies) assuming basic radial emission profiles (e.g. point source and S\'ersic) or allowing for arbitrary morphologies with input 2D brightness distributions.
GalSim is a widely used and adaptable package for producing mock galaxy images, which has been used to make detailed predictions for a range of upcoming instruments such as Euclid \citep[e.g.][]{EuclidXIII,EuclidXXVI}, Roman \citep[e.g.][]{Troxel2021,Troxel2023,Drakos2022} and the Rubin Observatory \citep[e.g.][]{Sanchez2020,LSST2021}. In this work, the CASTOR image simulator uses GalSim to generate mock CASTOR images based on current design specifications of the telescope. 

We have created the CASTOR image simulator to be easily customizable, with any exposure time. In this work we consider exposure times between 1000s and 360,000s, corresponding to the various planned CASTOR surveys. The image simulator is capable of generating images in the UV, UV-split, u, u-split, and g CASTOR filters. The images can be created either with or without dust attenuation and noise, although we include both in the images in this work.

Necessary input files include the point spread function (PSF) model and the bandpass files for the UV, UV-split, u, u-split, and g filters, provided by the CASTOR collaboration \citep{Cheng2023}. 
For the PSF model we use realistic models provided by CASTOR's primary Phase 0 contractor Honeywell Aerospace (private communication), 
which simulate multiple realisations of the PSF at different detector locations. We take the median PSF from the various realisations, and use the PSF from the centre of the detector, assuming that any spatial variations are negligible. We also convolve the model PSF with a Gaussian of $0\farcs023$ to account for spacecraft jitter.
The image simulator uses the key CASTOR imaging specifications valid as of the time of writing: 
1.0 m aperture diameter, pixel scale of $0\farcs1$/pixel, 1.0 e-/ADU gain, and a $0\fdg44$ by $0\fdg56$ (0.25 deg$^2$) field of view (FOV). These parameters can easily be modified in the script to simulate other telescopes. 

Four noise terms are applied to the generated images: sky noise, read noise, dark current, and shot noise.
The sky background, including a zodiacal, earthshine, and geocoronal emission line contribution, is computed from Hubble Space Telescope (HST) continuum data files \citep{pysynphot}. 
We implement the ``high'' earthshine model used in the HST ETC, defined as the earthshine at 38° from the limb \citep{STIShandbook}. 
For the zodiacal light we use the template used in the HST ETC from \citet{ZodiReport}, normalized to a Johnson V magnitude of 22.7 mag corresponding to the ``medium'' zodiacal level; we investigate how the assumed zodiacal light level affects our completeness limits in Appendix \ref{sec:noiseAppendix}.
Our sky-noise function outputs a single sky background flux value based on the integrated product of the chosen filter response function and this background (earthshine plus zodiacal) continuum spectrum. 
Airglow spikes as seen through the filter response are then added. 
We assume the ``average'' airglow estimate, with the [OIII] line at 2471\AA\ having an intensity of $1.50\times10^{–15}$ erg/s/cm$^2$/arcsec$^2$ \citep{STIShandbook}.

The sky background flux is converted from its output value $F_{\mathrm{sky[photons]}}$ in units of photons/s/cm$^2$/pixel to its value $F_{\mathrm{sky[ADU]}}$ in units of ADU/pixel using
\begin{equation} 
F_{\mathrm{sky[ADU]}} = F_{\mathrm{sky[photons]}} \pi g t_{\mathrm{exp}}\left(\frac{D}{2}\right)^2
\label{eq:skynoise_equation}
\end{equation}
where g is the gain in e-/ADU, $t_{\mathrm{exp}}$ is the exposure time in s, and D is the telescope aperture diameter in cm. 

The read noise is set to 3.0 e-/pixel, which is applied once per 3000s block of exposure time corresponding to approximately one CASTOR orbit. 
This is the best approximation at this stage, with the true read noise dependent on the decided readout strategy.
Both sky noise and read noise are added to the image using the GalSim \texttt{CCDNoise} function, which also adds in Poisson shot noise. 
Dark current of 0.002 e-/pixel/s is added separately using the GalSim \texttt{addNoise} function. 
This assumes an average of the dark current across the first two years of the mission when these surveys are expected to be conducted, as the dark current is predicted to vary linearly from 0.0001 e-/pixel/s at launch, to 0.01 e-/pixel/s after 5 years; we discuss these noise assumptions in further detail in Appendix \ref{sec:noiseAppendix}.
The images are output in units of ADU.

To create astrophysical images based on these telescope parameters, we require an object catalogue. In this work, we use simulated light-cones, which are described in Section \ref{sec:LightCones}.
The model that produces the light-cones assumes that each galaxy can be described by two S\'ersic profiles, representing a disk and bulge component. This is a much simpler model than, for example, in \citet{Marshall2022} and \citet{Fortuni2023} where hydrodynamical simulations are used to model realistic galaxy structure in their mock imaging. 
While GalSim could simulate images using more detailed input galaxy profiles, this is significantly more computationally expensive than assuming a basic profile, and would be prohibitive for simulating the very large CASTOR FOVs. Assuming basic S\'ersic profiles is a simple, robust, and reliable strategy for estimating the overall performance of a telescope.

\subsection{Simulated Light-Cones}
\label{sec:LightCones}

Our image simulator can take a light-cone catalogue as input and generate a corresponding field as seen by CASTOR. The input catalogue must contain the positions, S\'ersic properties, and fluxes of each object as required by GalSim.
In this work, we use a set of five realizations of 2-deg$^2$ lightcones constructed using the Santa Cruz semi-analytic model (SAM) for galaxy formation based on the same pipeline presented in \citet{2023MNRAS.519.1578Y}. The custom version of lightcones made for this work include additional photometry computed for CASTOR filters.
This simulated lightcone is $1\fdg4 \times 1\fdg4$ and contains galaxies over the redshift range $0 < z \lesssim 10$ and rest-frame UV magnitude range $-16 \gtrsim M_\text{UV} \gtrsim -25$ mag.

The dark matter halos in these lightcones are extracted from the SMDPL simulation from the MultiDark suite \citep{2016MNRAS.457.4340K}, which spans (400 Mpc h$^{-1}$)$^3$ and resolves halos down to $M_\text{h} \simeq 10^{10}$ M$_\odot$. These are projected onto a mock observed field using the \texttt{lightcone} package that is released as part of the \textsc{UniverseMachine} package \citep{2019MNRAS.488.3143B}. 
A total of five realizations of a 2-deg$^2$ field are generated by sampling different parts of the cosmological simulations, providing a total of $\sim 10$ deg$^2$ of simulated sky coverage. 
The dark matter halos in the lightcones preserve their relative distances and spatial distributions from the underlying cosmological simulation and provide redshift evolution along the line of sight that mimics the way we observe the real Universe. For halos in the lightcone, Monte Carlo merger trees are generated with the extended Press-Schechter formalism \citep[e.g.][]{1999MNRAS.305....1S} and the Santa Cruz SAM is used to track the evolution of galaxies therein.

The Santa Cruz SAM consists of a set of carefully curated recipes for galaxy formation and evolution that are either derived analytically or from observations and hydrodynamic simulations \citep{1999MNRAS.310.1087S, 2015MNRAS.453.4337S, 2021MNRAS.502.4858S}. These standard physical processes include gas cooling and accretion, multi-phase ISM gas partitioning, H$_2$-based star formation, stellar feedback, galaxy mergers, chemical evolution, black hole growth, and active galactic nuclei (AGN) feedback. 
The models have been shown to well-reproduce the observed one-point distribution functions of $M_\text{UV}$, $M_*$, and star formation rate \citep{2019MNRAS.483.2983Y, 2019MNRAS.490.2855Y}, observational constraints on the intergalactic medium (IGM) reionization history \citep{2020MNRAS.494.1002Y, 2020MNRAS.496.4574Y}, AGN populations $2 \lesssim z \lesssim 5$ \citep{2021MNRAS.508.2706Y}, as well as two-point auto-correlation functions from $0 < z \lesssim 7.5$ \citep{2022MNRAS.515.5416Y}.
We refer the reader to the schematic flow chart in \citet[][fig.~1]{2022MNRAS.515.5416Y} for a comprehensive illustration of the lightcone construction pipeline and the internal workflow of the Santa Cruz SAM, and Table A1 in \citet{2023MNRAS.519.1578Y} for the full set of quantities available in the mock lightcone. Here, we provide a concise summary of the model components that are most relevant to this work.

The Santa Cruz SAM tracks star formation in the \textit{disk} component and \textit{bulge} component of galaxies separately. When hot gas from cosmological accretion first cools, it is assumed to accrete into a disk, and stars that form out of that gas are assumed to have disk-like kinematics and morphology. disk stars can be moved into a bulge component via two mechanisms: galaxy-galaxy mergers and `disk instability'. For merging galaxy pairs with mass ratio larger than a critical value, all stars from both galaxies are placed into the bulge component of their descendant galaxy; and for mergers with lower mass ratio, the stars from the lower mass progenitor are deposited in the bulge component of the descendant galaxy. This approach is motivated by results from numerical simulations of galaxy mergers and is widely used in semi-analytic models. We refer the reader to \citet{2008MNRAS.391..481S, 2012MNRAS.426..237H, 2021MNRAS.508.2706Y} for the relevant references and further details.
The bulge component can also grow via a `disk instability' mode, which is triggered when the ratio of dark matter mass to disk mass falls below a critical value, causing the disk to become unstable. At each time-step, if the disk is unstable, a fraction of stars and gas is moved to the bulge such that the disk becomes marginally stable, with the cold gas assumed to trigger and be consumed by a starburst.

The full star formation and chemical enrichment histories of individual predicted galaxies are forward modelled into rest- and observed-frame photometry with SEDs generated based on the stellar population synthesis (SPS) models of \citet{2003MNRAS.344.1000B}. The observed-frame photometry is calculated accounting for dust attenuation effects using the attenuation curve of \citet{2000ApJ...533..682C}, and for absorption by hydrogen along the line of sight in the IGM \citep{1996MNRAS.283.1388M}.

S\'ersic profiles \citep{1963BAAA....6...41S, 1968adga.book.....S} of these simulated galaxies are determined based on the observed-frame luminosity of the disk and bulge components in the \textit{HST}/WFC3 F160W band, with S\'ersic indices and effective radii determined as described in \citet{2011ApJ...742...96W} and \citet{2015MNRAS.451.2933B}.

This catalogue includes five realizations of 2 deg$^2$ fields, each of which can contain six non-overlapping CASTOR pointings, making a total of 30 independent pointings. 
In this work we consider only one of the 2-deg$^2$ realizations, with 6 independent CASTOR pointings, which is a sufficiently large area for our study. 
The 2-deg$^2$ SAM light-cones provide a wide range of properties for each simulated object in the field, including but not limited to observable properties such as magnitudes in specific bands, redshift, celestial coordinates, and disk and bulge S\'ersic radii, and physical properties from the simulation such as average star formation rates, and stellar masses. 
From the light-cone catalogue, for each galaxy we use the properties: celestial coordinates (RA and Dec), intrinsic magnitude of the disk and bulge separately, combined dust-attenuated magnitude of both the disk and bulge, the disk S\'ersic radius, bulge radius, minor-to-major axis ratio as specified by $\cos{i}$ where $i$ is the inclination angle, and redshift. We randomly assign each galaxy with a position angle. We assume that the disk component of each galaxy has a S\'ersic index $n=1$, while bulges have $n=4$.

The CASTOR image simulator is capable of generating images with all of the redshifts in the lightcone ($z=0$--10), as we do throughout this work, or alternatively individual images per redshift slice for customisability. Based on a specified central RA and Dec for each image, objects within the FOV are included at their specified positions.

The 2-deg$^2$ SAM catalogue provides the total galaxy dust-attenuated magnitudes, however it does not provide dust-attenuated magnitudes for the galaxy disk and bulge separately.
Thus we estimate the separate dust-attenuated magnitudes by assuming that the bulge flux to total flux, and disk flux to total flux ratios, are equal for intrinsic and dust-attenuated fluxes:

\begin{equation} 
\mathrm{disk}_{D} = \mathrm{total}_{D} \frac{\mathrm{disk}_{ND}}{\mathrm{disk}_{ND}+\mathrm{bulge}_{ND}}
\label{eq:separating_dust_equation_1}
\end{equation}

\begin{equation} 
\mathrm{bulge}_{D} = \mathrm{total}_{D} \frac{\mathrm{bulge}_{ND}}{\mathrm{disk}_{ND}+\mathrm{bulge}_{ND}}
\label{eq:separating_dust_equation_2}
\end{equation}
where all variables are fluxes, the D subscript signifies `dust' flux values, and the ND subscript signifies `no dust' flux values. In reality, the bulge and disk components would have different levels of dust attenuation, however our simple assumption is the most reasonable estimate given the information that is available. As we only use the SAM to understand basic source detectability in this work, we do not expect that this approximation will have any significant impact on our results. 

Catalogue magnitudes are converted to flux densities in $\mu$Jy using Equation \ref{eq:mag_2_flux}.
\begin{equation} 
\mathrm{flux \: density} = 10^{{-\frac{(m - 23.9)}{2.5}}}
\label{eq:mag_2_flux}
\end{equation}
The flux density is converted to a flux in photons/s/cm$^2$ by integrating the product of the chosen filter response and the flux density over the spectral range. The flux is then converted to ADU/pixel using Equation \ref{eq:skynoise_equation}. 

The bulge and disk radii in kpc are converted to arcseconds based on the galaxy's redshift and our assumed cosmology. The galaxy object is generated by combining separate disk and bulge S\'ersic profiles from the catalogue S\'ersic radii for each component, using GalSim's \texttt{Sersic} function. Observed galaxy shapes can be simulated using various parameters in the GalSim \texttt{.shear} function. This is currently specified by a randomly generated position angle $\beta$ (from a uniform distribution) and the axis ratio `cos$i$' given in the catalogue (or `q' in GalSim). 

Each galaxy is convolved with the CASTOR PSF for the chosen filter using the GalSim function \texttt{Convolve}. The single galaxy image is then added to the full field image at the position corresponding to the galaxy's coordinates. After cycling through all galaxies in the catalogue, noise is added to the full field image as described in Section \ref{sec:GalSim}. The image is saved as a FITS file. Using a 2 deg$^2$ field from the SAM catalogue, each full field image contains over 1.5 million galaxies.

\section{Images}
\label{sec:Images}

\begin{figure*}
\begin{center}
\begin{adjustwidth}{-0.125\textwidth}{-0.125\textwidth}
\begin{center}
\includegraphics[width=0.274\linewidth]{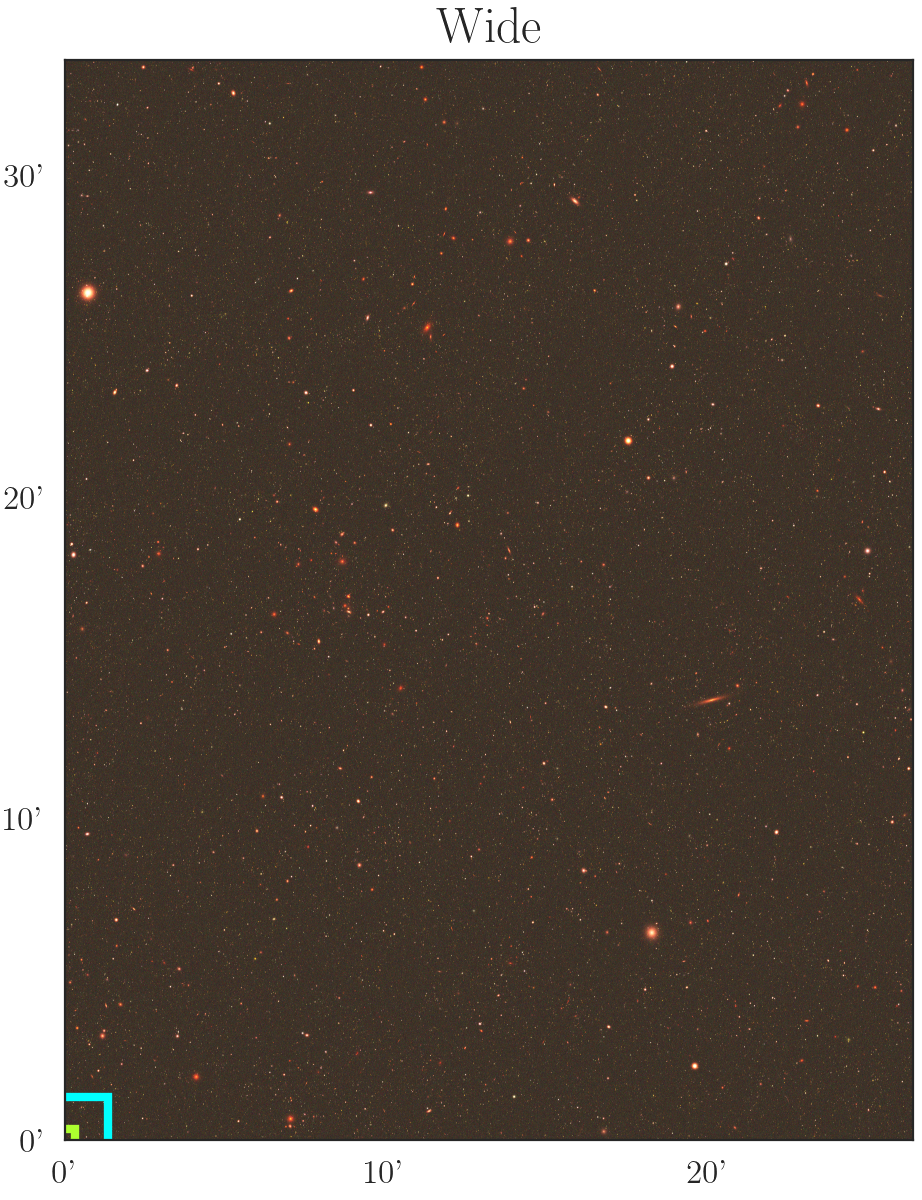}
\hspace{0.07cm}
\includegraphics[width=0.273\linewidth]{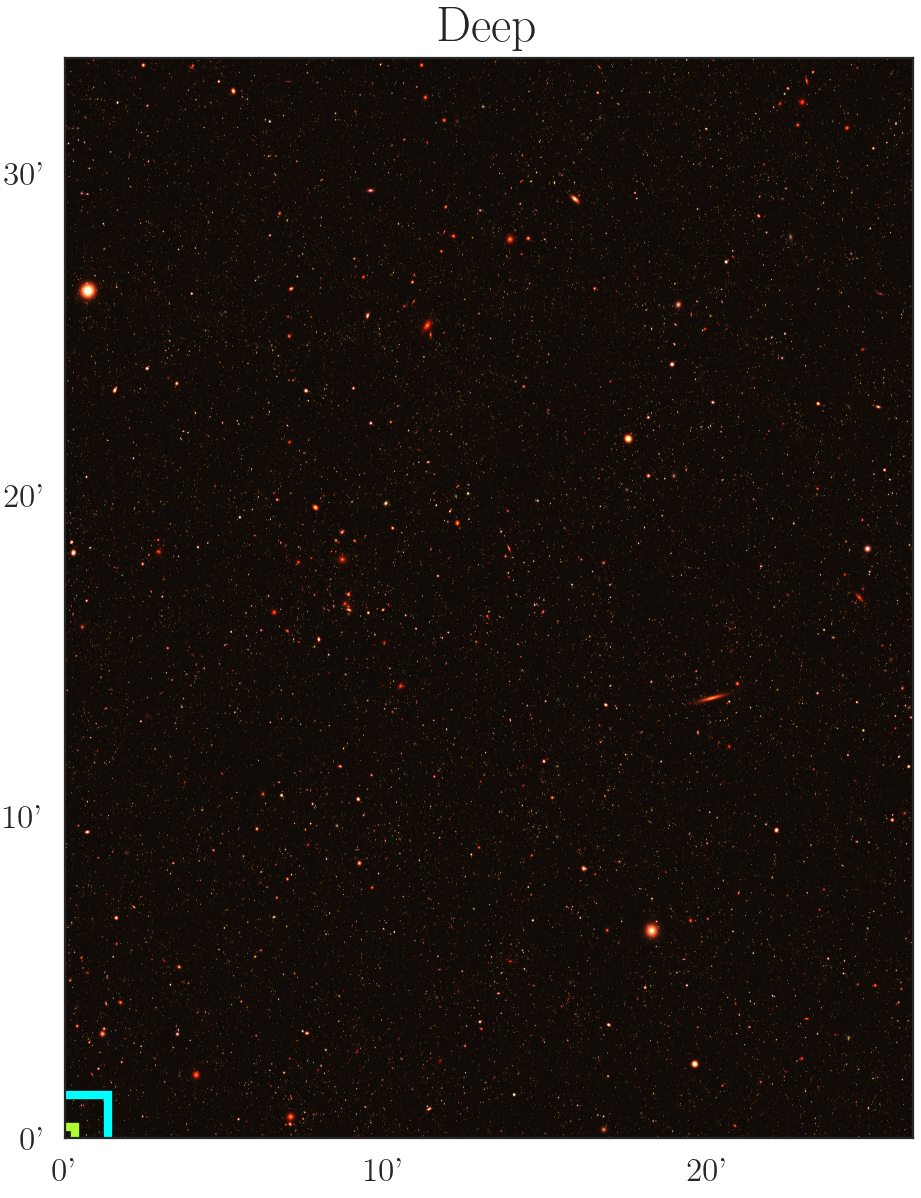}
\hspace{0.08cm}
\includegraphics[width=0.274\linewidth]{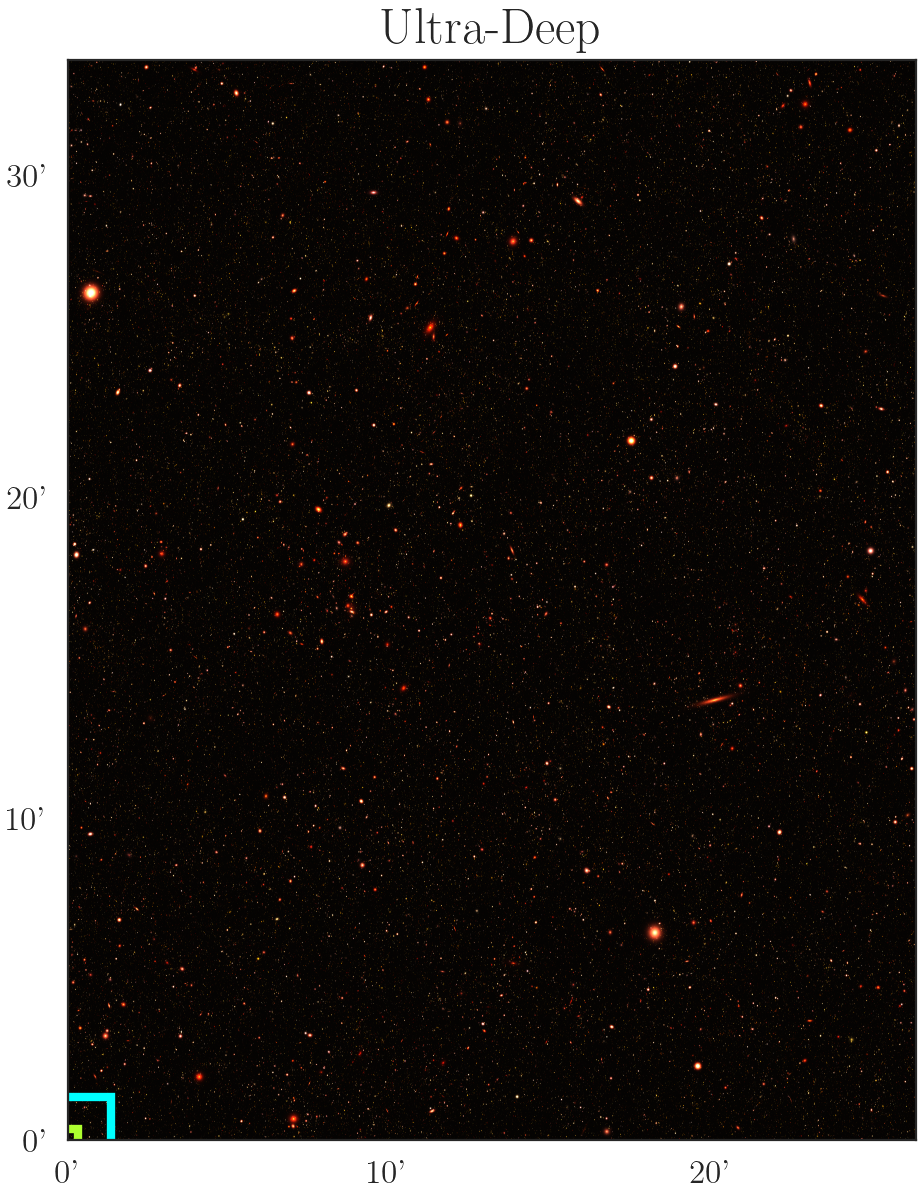}
\hspace{0.08cm}

\includegraphics[width=0.28\linewidth]{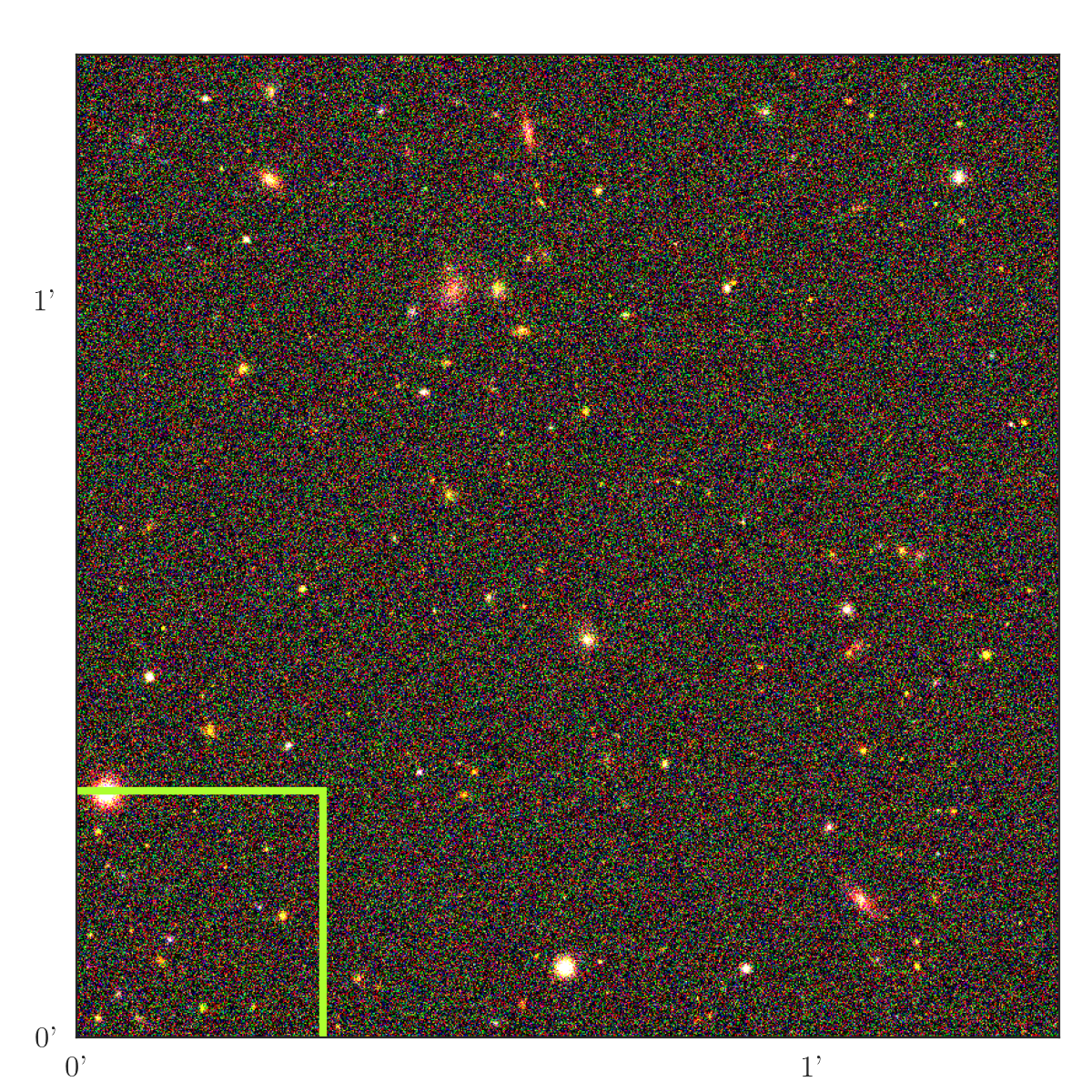}
\includegraphics[width=0.28\linewidth]{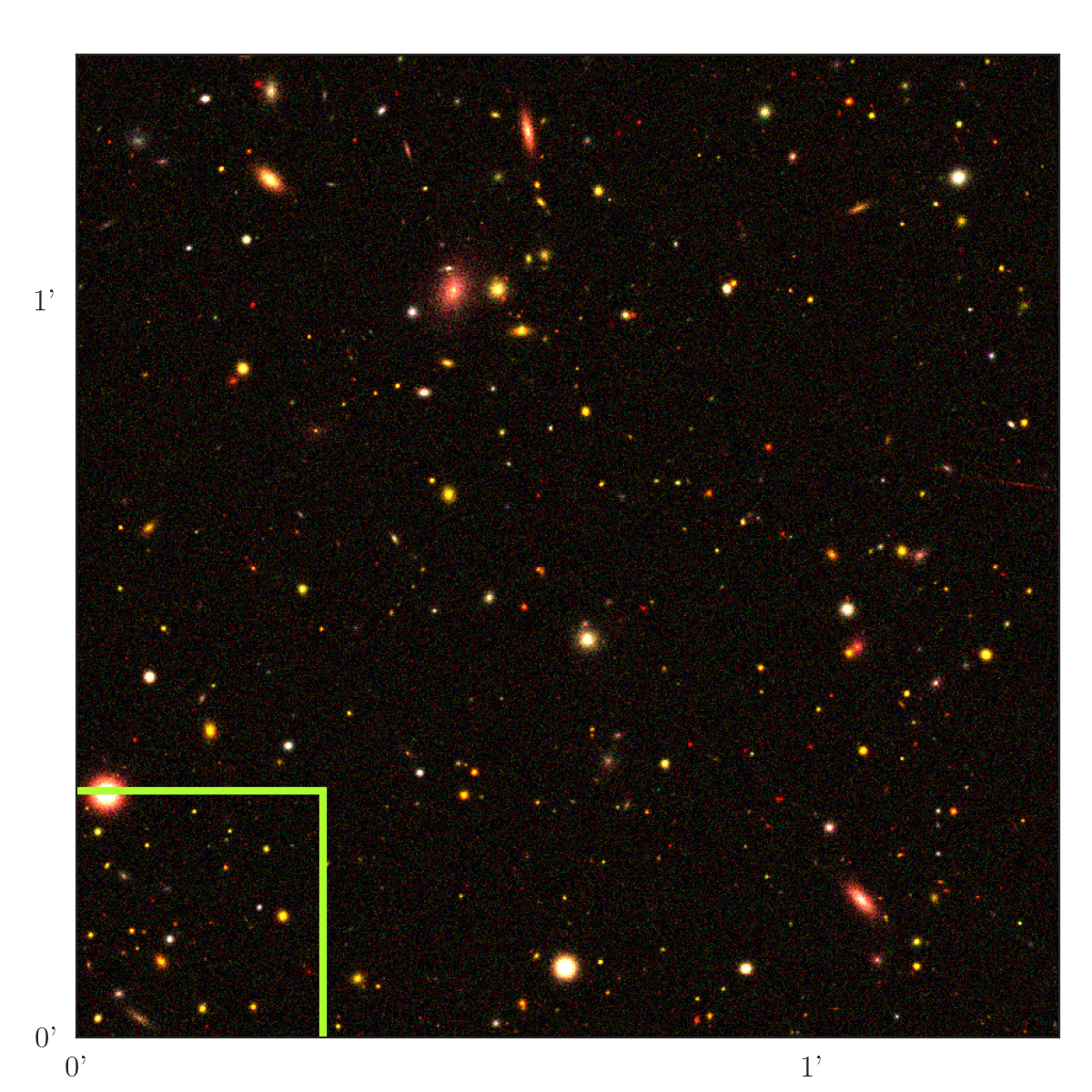}
\includegraphics[width=0.28\linewidth]{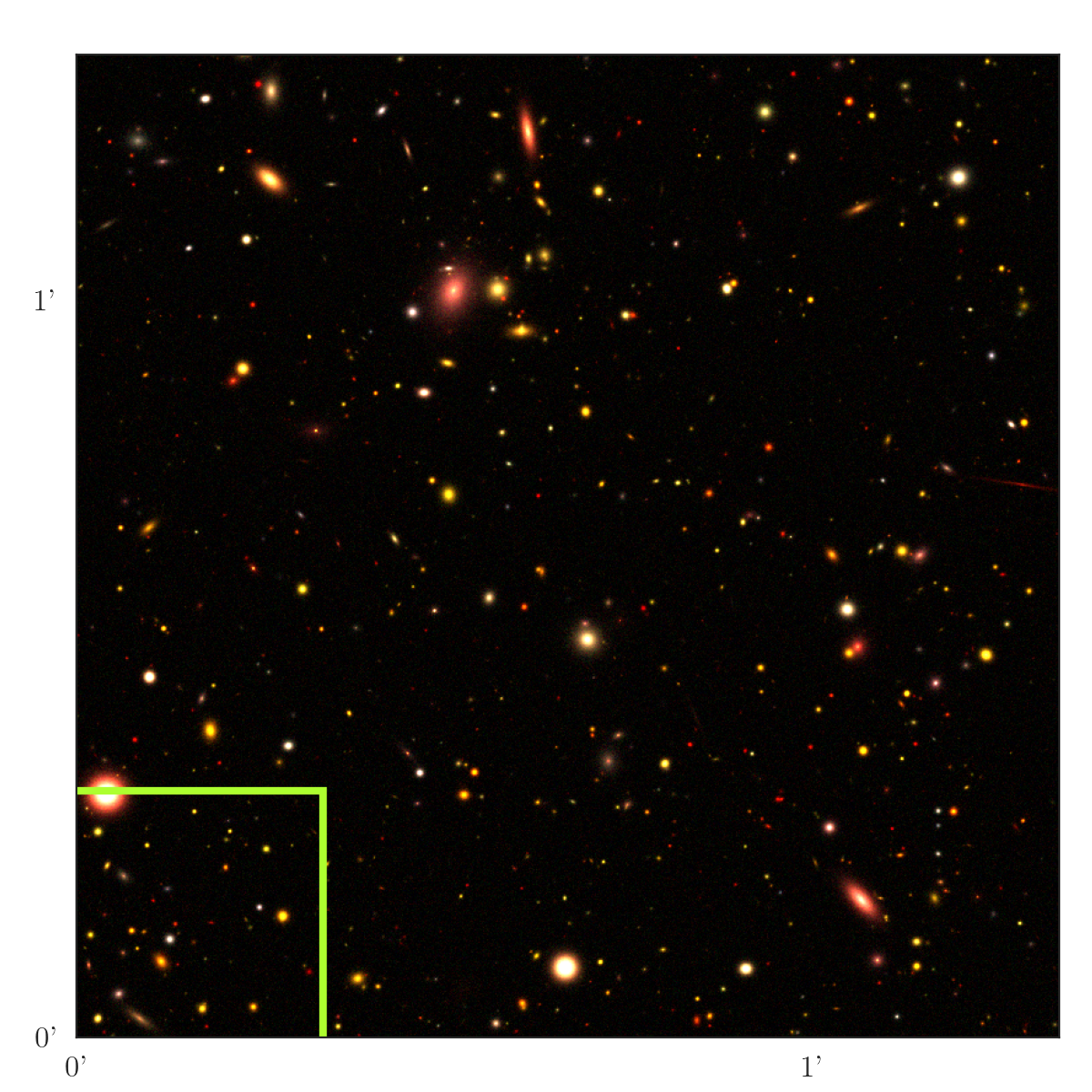}

\includegraphics[width=0.28\linewidth]{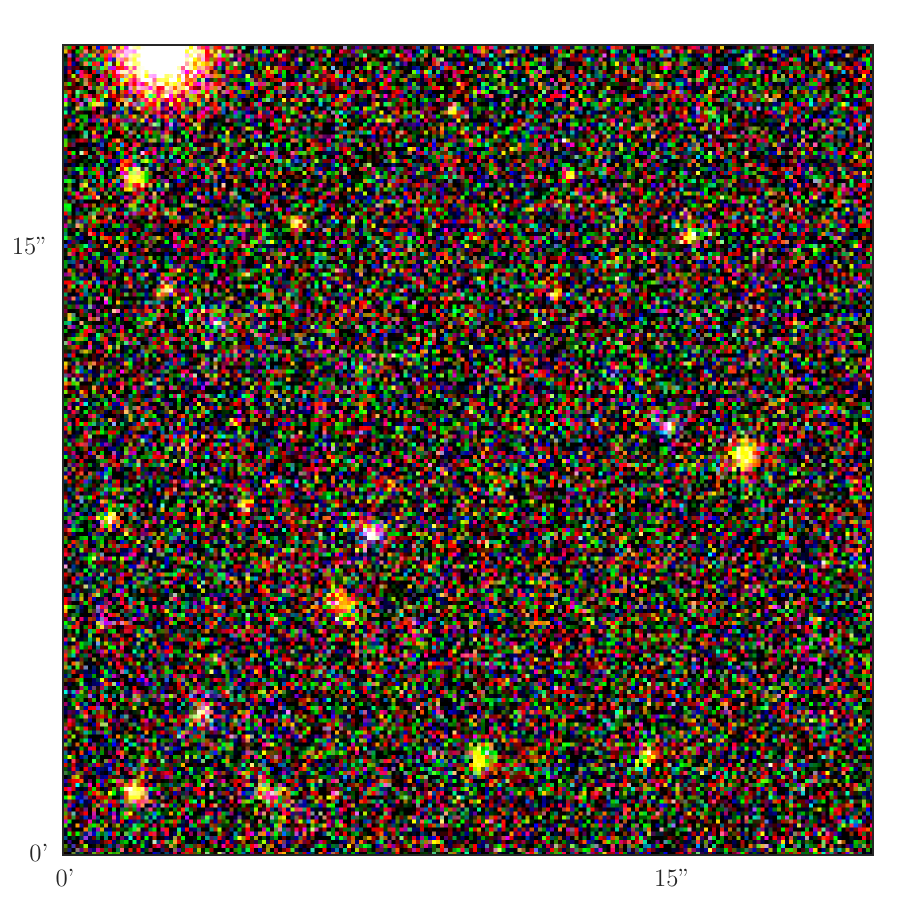}
\includegraphics[width=0.28\linewidth]{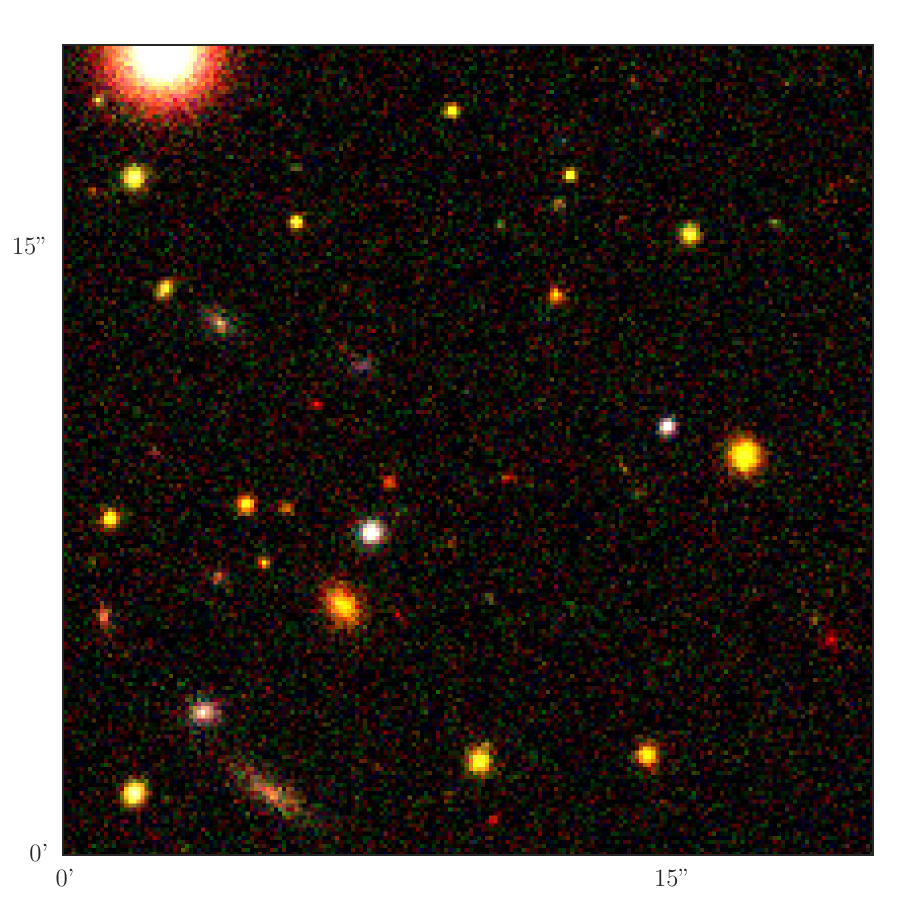}
\includegraphics[width=0.28\linewidth]{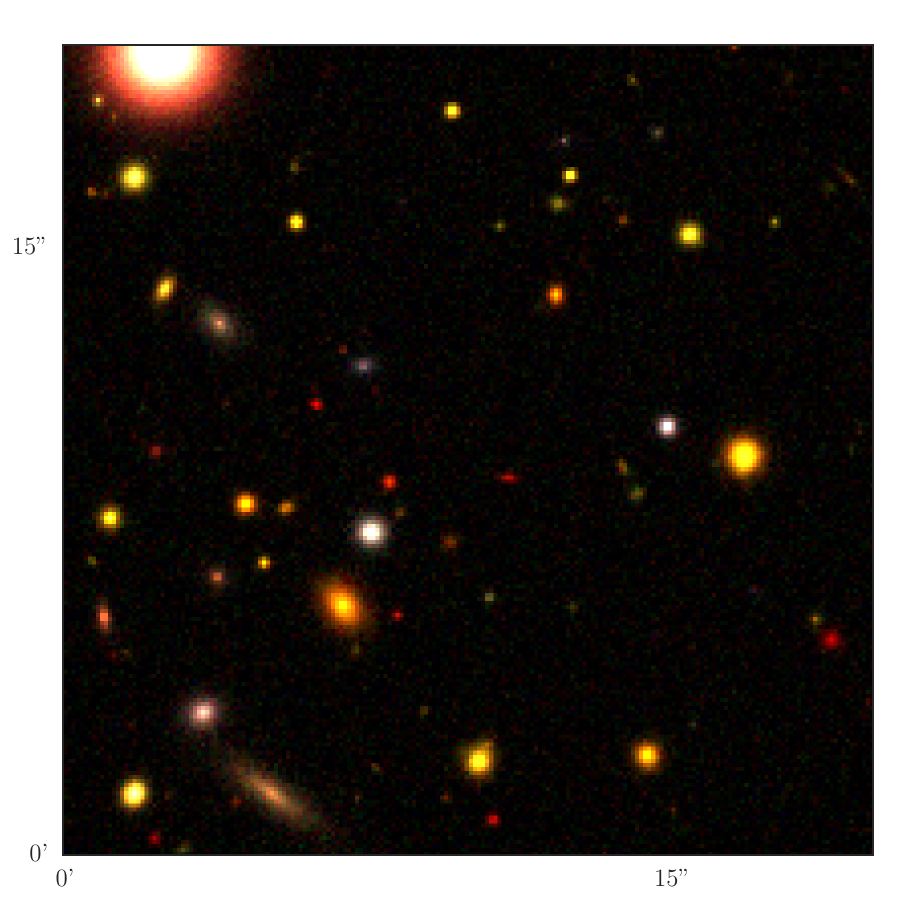}
\end{center}
\end{adjustwidth}
\caption{Example mock CASTOR images for the three surveys: Wide (left), Deep (middle) and Ultra-Deep (right). The upper panels show the full CASTOR FOV, with the lower panels zooming in to show more detail. The blue and green squares depict the zoom-in regions for the middle and bottom panels, respectively, for perspective.
These RGB images are made using the g, u and UV bands, with the images scaled by the exposure time (i.e. in ADU/s) to account for the g-band exposure times being twice as long. Each image has the same normalization and the same arcsinh scaling, for consistency. We note that these images are made only using a galaxy catalogue, and include no foreground stars.}
\label{fig:RBG}
\end{center}
\end{figure*}

\begin{figure*}
\begin{center}
\includegraphics[scale=0.62]{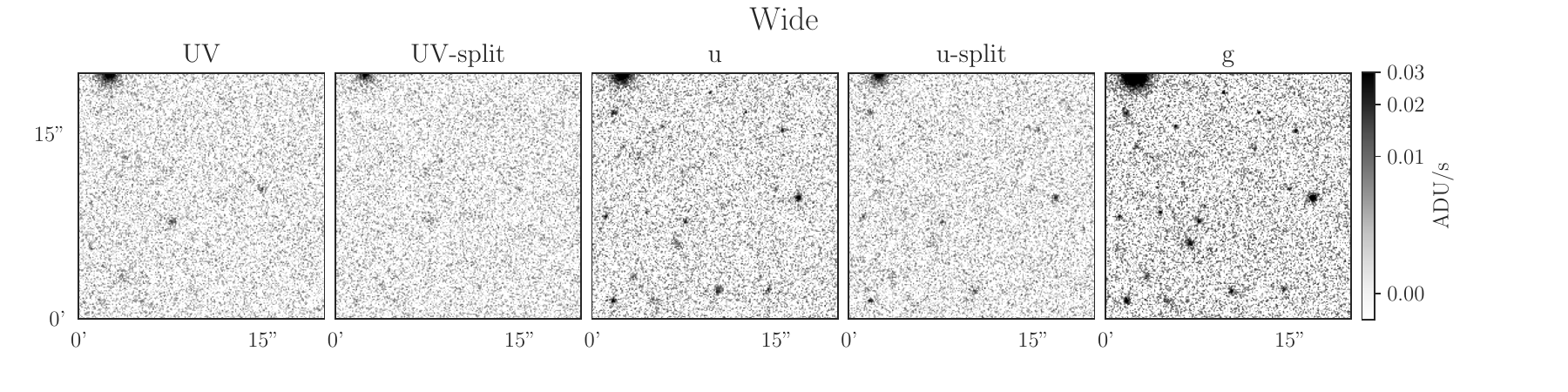}
\includegraphics[scale=0.62]{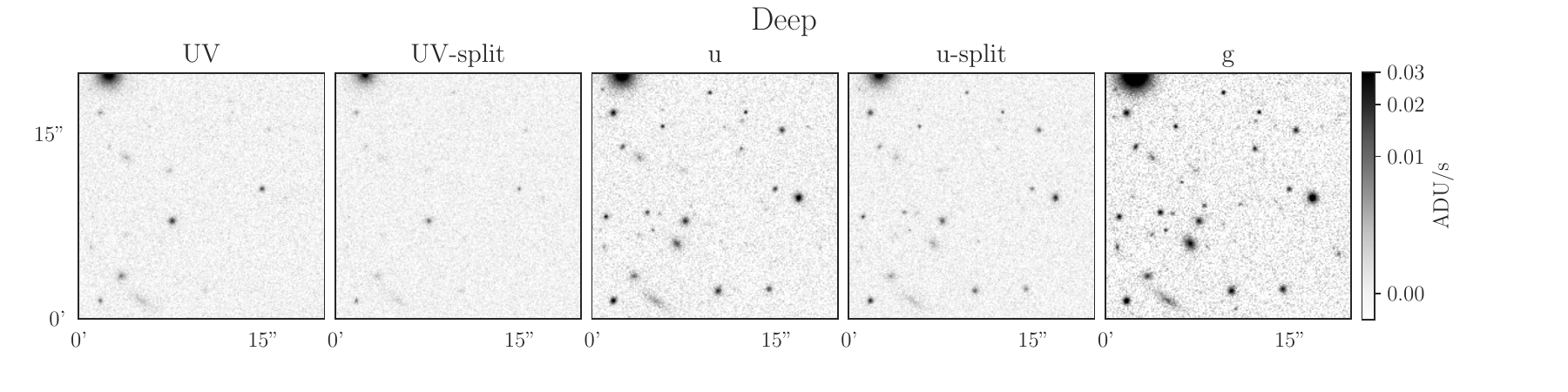}
\includegraphics[scale=0.62]{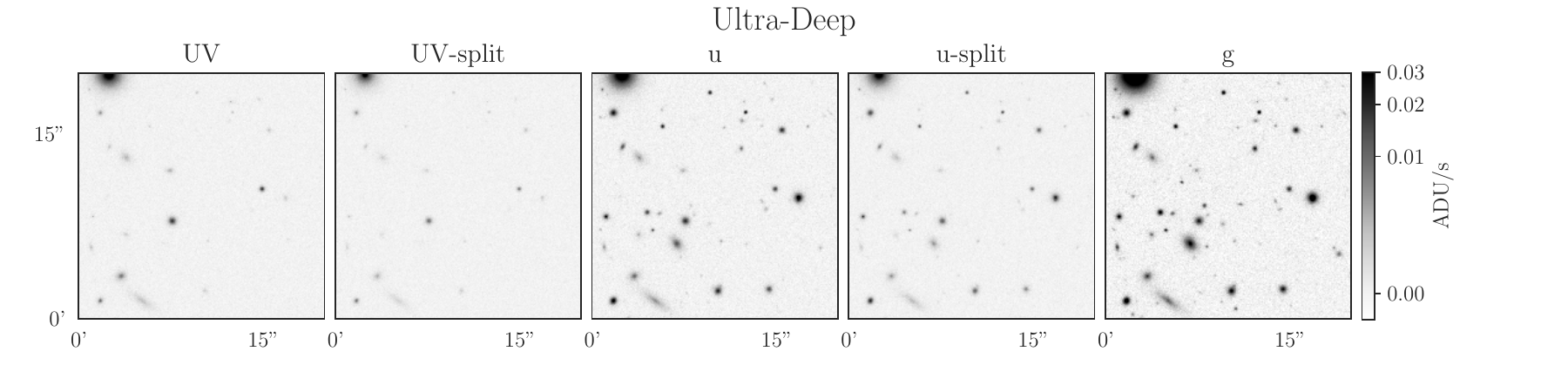}
\caption{Example mock CASTOR images for the three surveys: Wide (top), Deep (middle) and Ultra-Deep (bottom), in each of the five CASTOR filters. These are of a smaller FOV than a full CASTOR image, zoomed-in to the yellow square region depicted in Figure \ref{fig:RBG}, equivalent to the bottom rows of that figure.
We note that the exposure time in the g-band images is twice that of the other filters, due to the telescope and survey design.
The images are scaled by the exposure time (i.e. in ADU/s) to account for the longer g-band exposures. 
Each image has the same normalization and the same arcsinh scaling, for consistency. }
\label{fig:allFilts}

\end{center}
\end{figure*}

\subsection{CASTOR survey images}

\begin{figure*}
\begin{center}

\includegraphics[scale=0.62]{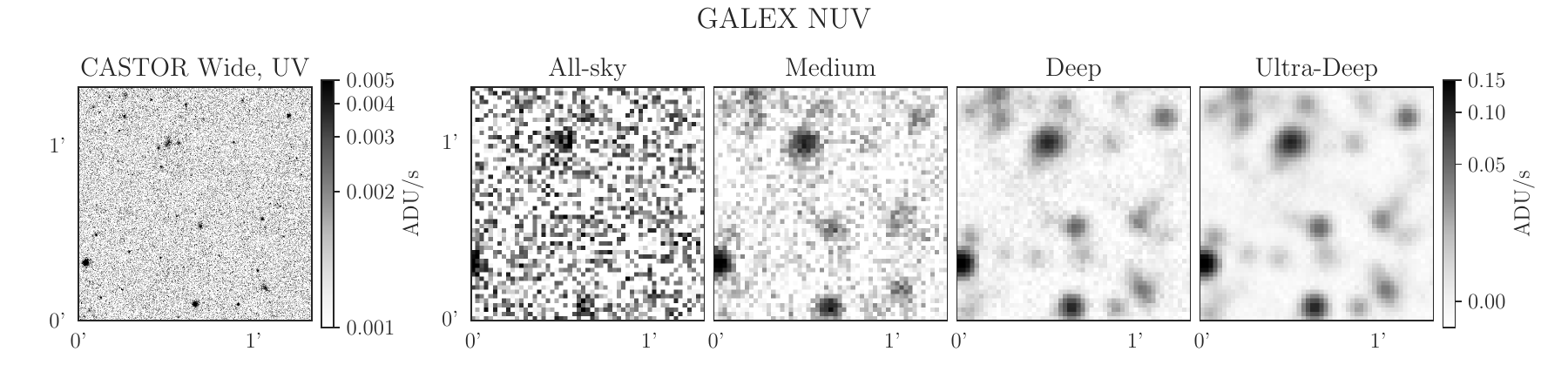}
\vspace{-0.3cm}
\includegraphics[scale=0.62]{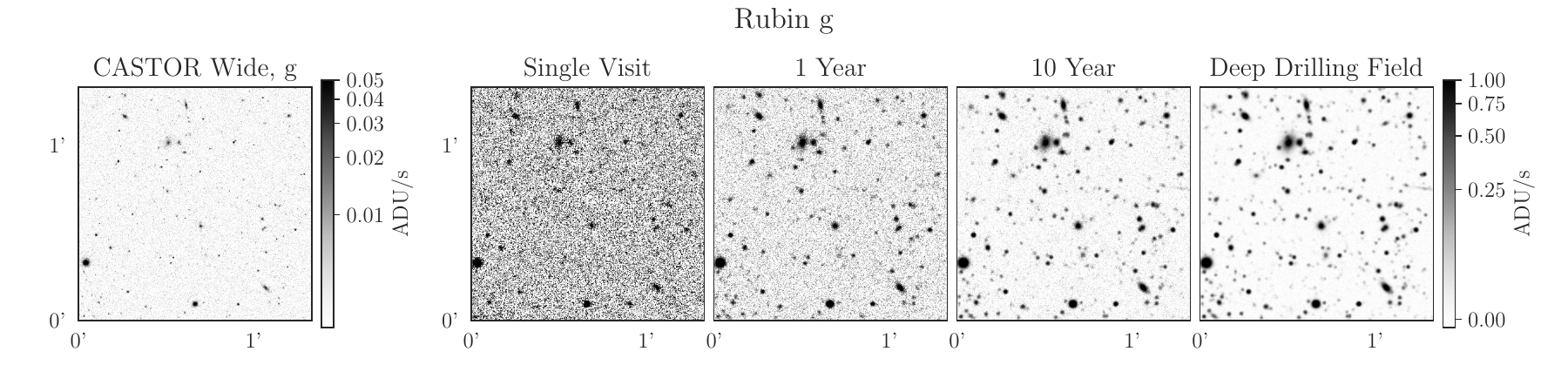}
\vspace{-0.3cm}
\includegraphics[scale=0.62]{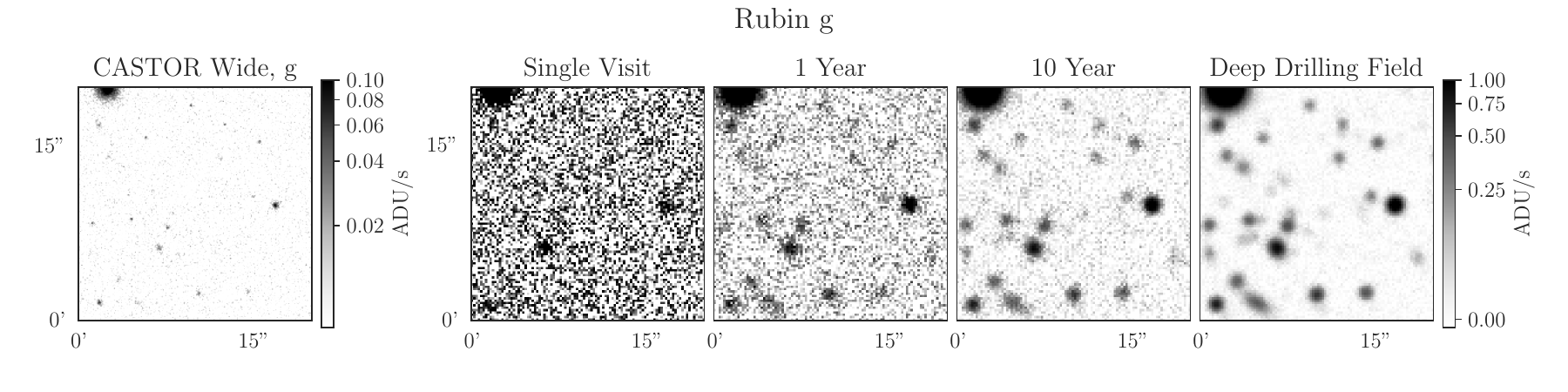}
\vspace{-0.3cm}
\caption{Example mock CASTOR images for the Wide survey, compared to simulated GALEX and Rubin images of the same field. 
To compare to GALEX we use the CASTOR UV and GALEX NUV filters.
We show GALEX exposure times corresponding to those from the All-sky (0.1ks), Medium (1.5ks), Deep (30ks), and Ultra-Deep (200ks) Imaging Surveys.
To compare to Rubin, we use the CASTOR and Rubin g filters.
For the four Rubin panels, we simulate a single visit exposure (30s), the 1 year depth (300s), the 10 year depth (3000s), and the
final depth in each of Rubin’s four Deep Drilling Fields (39000s).
We note that these are of a smaller FOV than a full CASTOR image, chosen to highlight the differences in the images from these observatories.
The top two panels show a zoom of $\sim1'$ per side, corresponding to the red square region depicted in Figure \ref{fig:RBG}, equivalent to the middle row of that figure. 
The bottom panels show a closer zoom-in of this region, corresponding to the yellow square region depicted in Figure \ref{fig:RBG}, equivalent to the bottom row of that figure.
}
\label{fig:GALEX_LSST}

\end{center}
\end{figure*}

Here we present example CASTOR images created using our image simulator, for each of the planned CASTOR surveys: Wide, Deep, and Ultra-Deep. For each exposure time and filter combination, we create six images of various CASTOR FOVs within the 2-deg$^2$ SAM catalogue. This allows us to test our codes and predictions with more galaxies, as well as giving estimates for the effect of cosmic variance on the number of galaxies expected in each image. In this section we show only one of these FOVs as an example, although we use all of these in our quantitative analysis in Section \ref{sec:Results}. We note that our images are made only using a galaxy catalogue, and include no foreground stars. Thus we cannot investigate issues such as star--galaxy blending, where some of the detectable galaxies in our images would, in reality, be obscured and undetectable because of overlapping foreground stars. However, this allows us to more easily extract sources and perform a comparison to our input galaxy catalogue, which is adequate for our galaxy-focused study. With the high resolution of CASTOR, we expect that this will have a negligible impact on the completeness limits that we estimate in this work.

In Figure \ref{fig:RBG} we show colour images of a single pointing of the three CASTOR surveys, created by stacking the g, u and UV images. These show the improvement in the detail that can be extracted from the progressively deeper surveys. While the Wide survey will cover the largest area on the sky, it is clear that the properties of the galaxies within the images, particularly the smallest and faintest galaxies, will be more difficult to extract. The Ultra-Deep survey, in contrast, will cover a small area on the sky, but with images showing exquisite detail. As expected, the deeper surveys will discover fainter galaxies, which we investigate quantitatively in Section \ref{sec:Results}.

In Figure \ref{fig:allFilts} we show example images for the planned surveys from each of the filters separately.
We can see that the split filters, UV-split and u-split, observe less flux than their un-split counterparts, as expected due to their reduced bandwidth. Generally we see the lowest flux in the UV and UV-split filters, due to the low throughput in these bandpasses. The g-band images have twice the exposure time as the other filters, hence their significantly improved depth.

\subsection{Comparison to GALEX and Rubin}
To understand the capability of CASTOR in context of past and future observatories, we use the same image generation software to create mock images for both the Galaxy Evolution Explorer (GALEX) and the Rubin observatory.

GALEX was an orbiting space telescope in operation from 2003--2013, which took images in two bands, the near-UV (NUV) and far-UV, as well as low-resolution grism spectroscopy.
The mission's primary aim was an investigation of the evolution of star formation in the Universe, providing the first all-sky UV survey \citep{Martin2005}.
However, with an angular resolution of $\sim5''$, GALEX could only detect the most massive galaxies out to $z<2$, and resolve star formation within only the most nearby systems. CASTOR is designed to offer a significant improvement in  sensitivity and resolution compared to GALEX.

To simulate GALEX images for a comparison to CASTOR, we use its telescope parameters diameter $D = 0.5$m, pixel scale $1\farcs5$/pixel, a gain of 1.0 e-/ADU, with no dark current or read noise.
We simulate images in the NUV filter, assuming a PSF FWHM of $6\farcs0$, which we assume is a Gaussian in shape \citep{Morrissey2005}, and a sky background level of 26.5 mag/arcsec$^2$ \citep{Martin2005}.
For exposure times, we use those from the All-sky Imaging Survey (AIS), 0.1ks, Medium Imaging Survey (MIS), 1.5ks, Deep Imaging Survey (DIS),  30ks, and the
Ultra-Deep Imaging Survey (UDIS), 200ks
\citep{Martin2005}. The resulting images are shown in Figure \ref{fig:GALEX_LSST}. These clearly show the improvement in UV-imaging that CASTOR will have over GALEX, with significantly improved depth and resolution. The GALEX resolution was $\sim6\farcs0$ in the NUV, compared to CASTOR's planned $\sim0\farcs15$. This $\sim$30-fold improvement in resolution will mitigate the problem of source blending in GALEX imaging, and significantly improve our ability to identify UV sources and accurately measure their properties --- both in the local universe and across cosmic timescales \citep{Martin2005,Noeske2007,Salim2007,GildePaz2007,Thilker2007}.

The Rubin Observatory’s primary survey, the LSST, will be a photometric imaging survey covering 18,000 square degrees to image billions of galaxies over hundreds of repeat observations in the optical \citep{Ivezic2019}. Survey operations are planned to begin in 2025. LSST will cover four main science topics: probing dark energy and dark matter, exploration of the solar system and the transient optical sky, and mapping the Milky Way. Many of these science topics overlap with CASTOR goals. Therefore, CASTOR data can contribute in a unique way to multiple key topics with LSST. 

To make simulated Rubin images, we use a diameter $D = 6.423$m, a pixel scale of $0\farcs2$/pixel, dark current of 0.02 e-/pixel/s, a read noise of 9 e-/pixel, and a gain of 1.0 e-/ADU \citep{OConnor2019}.
We simulate images in the g-band, with a sky background of 22.26 mag/arcsec$^2$ \citep{Ivezic2019}.
We assume that the PSF is the convolution of two Gausssians, one with FWHM $0\farcs8$ and one with $0\farcs3$, each with the same flux, to be consistent with Imsim, the LSST simulation software package\footnote{\url{https://lsstdesc.org/imSim}}.
We simulate exposure times of 30s, corresponding to a single visit, 300s, the 1 year depth, 3000s, the 10 year depth, and 39000s, the final depth in each of LSST’s four Deep Drilling Fields (see C\^ot\'e et al. 2024, in preparation). The resulting images are shown in Figure \ref{fig:GALEX_LSST}. While the improvement of the CASTOR images is not as dramatic as for GALEX, the zoomed-in panels still clearly show the effect of the increased resolution of CASTOR over LSST. With LSST's FWHM$\simeq0\farcs8$ due to atmospheric seeing, the space-based CASTOR offers a factor of $\sim5$ improvement in spatial resolution.

LSST will provide an unprecedented volume of data with high constraining power. Beyond the volume and the processing of the data, many challenges could limit its future science measurements. For example, one systematic uncertainty will be the potential blending of sources in LSST deep field imaging. The other challenge that must be addressed is the accurate estimation of photometric redshifts (photo-zs) due to the lack of spectroscopic resources compared to the volume of the data \citep{Ansari.2019}. The challenges mentioned are critical for the future studies of galaxies with LSST data  and are extended to other LSST science topics including weak lensing and Type Ia Supernovae.

The combination of a high angular resolution and a sharp PSF will enable CASTOR to highly contribute to resolving the blends in LSST imaging, reducing the future systematic measurement uncertainties. It will also provide robust measurements of galaxy shapes, essential for weak lensing detection and the study of galaxy evolution. The angular resolution of CASTOR will also help to improve strong lensing modeling for the systems observed by LSST, and CASTOR’s UV/blue-optical data will contribute to enhance the SED fitting for lens modeling. Furthermore, \citet{Graham_2020} showed that extending the wavelength coverage of LSST photometric data with CASTOR UV and u-band photometry could reduce the standard deviation of galaxy photo-z estimates by $\simeq~30~\%$ and the fraction of outliers by $\simeq~50~\%$.

These images show examples of the expected image quality reached with each of the CASTOR surveys. In the following section, we use these images to study quantitatively the expected depth and completeness limits of these surveys.

\section{Results}
\label{sec:Results}

\subsection{Source Extraction}
We now use our simulated images to estimate the completeness of the various key CASTOR surveys.

We extract the sources in each of our mock images using Source Extractor \citep{SExtractor}.
We extract galaxies with at least 4 contiguous pixels above 2$\sigma$. In these mock galaxy images we do not have concerns of stars, cosmic rays or artefacts being incorrectly identified as galaxies. Thus our choice in extraction parameters is optimized to detect small, faint objects, and we do not need to perform any star--galaxy classifications that would be otherwise required in true images.

From each of the input source positions, we find the nearest extracted source. If the matched source is within $0\farcs3$, we consider the input source as having been successfully detected. We do not include sources in our detected or input number counts that have multiple matches within $0\farcs3$, for simplicity.
To reduce computational expense, we only include input galaxies with magnitude $\leq31$ mag in our search list, as from our point source analysis we find that no galaxies will be detected beyond this threshold (see below).
This matching allows us to determine which of the input galaxies are detected, and thus analyse the completeness as functions of various input parameters. This also allows us to compare the input and measured properties to determine the accuracy of our CASTOR measurements.

\subsection{Magnitude Completeness}

\begin{table}
\caption{Magnitude completeness limits for each filter in each survey. These limits vary for point sources (top panels) and galaxies (bottom panels). The faintest detectable source has been detected with our Source Extractor configuration, which requires at least 4 pixels with $\rm{SNR}>2$.}
\label{tab:CompletenessLims}
\begin{tabular}{llllll}
\hline 
\hline 
Survey
& UV
& UV-split
& u
& u-split
& g
\\
& & & (mag) &  & \\
\hline
\multicolumn{5}{l}{75\% Completeness for Point Sources} \\
\hline
Wide
& 27.2
& 26.2
& 27.0
& 26.4
& 26.9
\\
Deep
& 28.8
& 28.2
& 28.5
& 28.4
& 28.4
\\
Ultra-Deep
& 30.2
& 29.5
& 29.7
& 29.6
& 29.6
\\
\hline
\multicolumn{5}{l}{Faintest Detectable Point Source} \\
\hline
Wide
& 27.7
& 26.8
& 27.6
& 27.2
& 27.4
\\
Deep
& 29.5
& 28.7
& 29.4
& 29.3
& 29.3
\\
Ultra-Deep
& 30.9
& 29.9
& 30.8
& 30.5
& 30.5

\\

\hline
\multicolumn{6}{l}{75\% Completeness for Galaxies ($0<z<5$)} \\
\hline

Wide & 24.8 & 22.8 & 24.9 & 24.0 & 25.0 \\
Deep & 27.6 & 26.7 & 27.6 & 27.2 & 27.3 \\
Ultra-Deep & 29.2 & 28.3 & 28.9 & 28.7 & 28.7\\

\hline
\multicolumn{5}{l}{Faintest Detectable Galaxy} \\
\hline

Wide &27.5 & 26.6 & 27.4 & 26.9 & 27.2 \\
Deep &29.4 & 28.6 & 29.2 & 28.9 & 29.0 \\
Ultra-Deep & 30.7 & 29.9 & 30.5 & 30.2 & 30.3

\\

\hline
\end{tabular} 

\end{table}

\begin{table*}
\caption{The number of galaxies detected in the various CASTOR surveys, as predicted by our mock images. We have simulated 6 CASTOR FOVs, with exposure times equivalent to each survey. The first set of columns shows the mean number of galaxies detected in these 6 pointings, with errors showing the range of the sparsest and most dense FOVs.
The second set of columns shows our predictions for the full survey areas. We multiply our expected range in one pointing by the survey area: 2227 deg$^2$ for Wide, 83 deg$^2$ for Deep, and 1 deg$^2$ for Ultra-Deep, or 8909, 332, and 4 pointings respectively. This assumes that our 6 FOVs reasonably sample the expected cosmic variance.
}
\label{tab:GalaxyCounts}
\begin{tabular}{lrrrrrr}
\hline 
\hline 
& \multicolumn{3}{c}{Number of Galaxies in 1 Pointing} & \multicolumn{3}{c}{Number of Galaxies in Full Survey} \\
&($\times 10^{3}$) & ($\times 10^{3}$) & ($\times 10^{3}$)
& ($\times 10^7$)
& ($\times 10^7$)
& ($\times 10^5$)
\\

\hline 
Filter & Wide & Deep & Ultra-Deep & Wide & Deep & Ultra-Deep \\
\hline 

UV & $11.6 \substack{ +0.4\\-0.3 }$
& $58.0 \substack{ +1.5\\-1.2 }$
& $121.4 \substack{ +3.9\\-2.3 }$
& $10.3 \substack{ +0.3\\-0.2 }$
& $1.9 \substack{ +0.0\\-0.0 }$
& $4.9 \substack{ +0.2\\-0.1 }$
\\
UV-split & $5.9 \substack{ +0.2\\-0.2 }$
& $45.5 \substack{ +1.5\\-1.0 }$
& $104.4 \substack{ +3.8\\-2.4 }$
& $5.3 \substack{ +0.2\\-0.2 }$
& $1.5 \substack{ +0.0\\-0.0 }$
& $4.2 \substack{ +0.2\\-0.1 }$
\\
u & $28.7 \substack{ +1.3\\-1.1 }$
& $124.2 \substack{ +5.0\\-4.7 }$
& $255.0 \substack{ +9.5\\-6.6 }$
& $25.6 \substack{ +1.1\\-1.0 }$
& $4.1 \substack{ +0.2\\-0.2 }$
& $10.2 \substack{ +0.4\\-0.3 }$
\\
u-split & $16.3 \substack{ +0.8\\-0.5 }$
& $94.3 \substack{ +4.3\\-4.6 }$
& $200.7 \substack{ +8.3\\-7.3 }$
& $14.5 \substack{ +0.7\\-0.4 }$
& $3.1 \substack{ +0.1\\-0.2 }$
& $8.0 \substack{ +0.3\\-0.3 }$
\\
g & $43.2 \substack{ +1.8\\-1.3 }$
& $161.9 \substack{ +4.5\\-3.8 }$
& $336.4 \substack{ +7.4\\-5.2 }$
& $38.5 \substack{ +1.6\\-1.2 }$
& $5.4 \substack{ +0.1\\-0.1 }$
& $13.5 \substack{ +0.3\\-0.2 }$
\\

\hline 
\end{tabular} 
\end{table*}

\subsubsection{Point Sources}
\label{sec:pointsources}

We first test our imaging and magnitude extraction techniques by simulating images containing only point sources. In the CASTOR FOV we add 10,000 point sources, with magnitudes sampled from a uniform distribution from 20--33 mag (with the same magnitude in each filter), and randomly selected locations. We extract the sources from these images equivalently to the galaxy images, so that we can compare the point source limit to the completeness for extended galaxies.

In Table \ref{tab:CompletenessLims} we list the 75\% point source completeness limits for the Wide, Deep and Ultra-Deep surveys in each of the five filters.
We predict that the Wide survey will be 75\% complete down to 26.2--27.2 mag, with the Deep survey extending to 28.2--28.8 mag, and the Ultra-Deep survey reaching to 29.5--30.2 mag. 
Table \ref{tab:CompletenessLims} also lists the faintest point source detected in the images: 26.8--27.7 mag for Wide, 28.7--29.5 mag for Deep, and 29.9--30.9 mag for Ultra-Deep.
The UV filter is predicted to have the best sensitivity in all surveys, while the UV-split filter is the shallowest. However, all filters are generally well matched in depth for each survey.
These are very similar to the estimated 5-sigma point source depths predicted by the CASTOR ETC \citep{Cheng2023}.

\subsubsection{Galaxies}

\begin{figure*}
\hspace{-1cm}
\vspace{-1cm}
\includegraphics[scale=0.8]{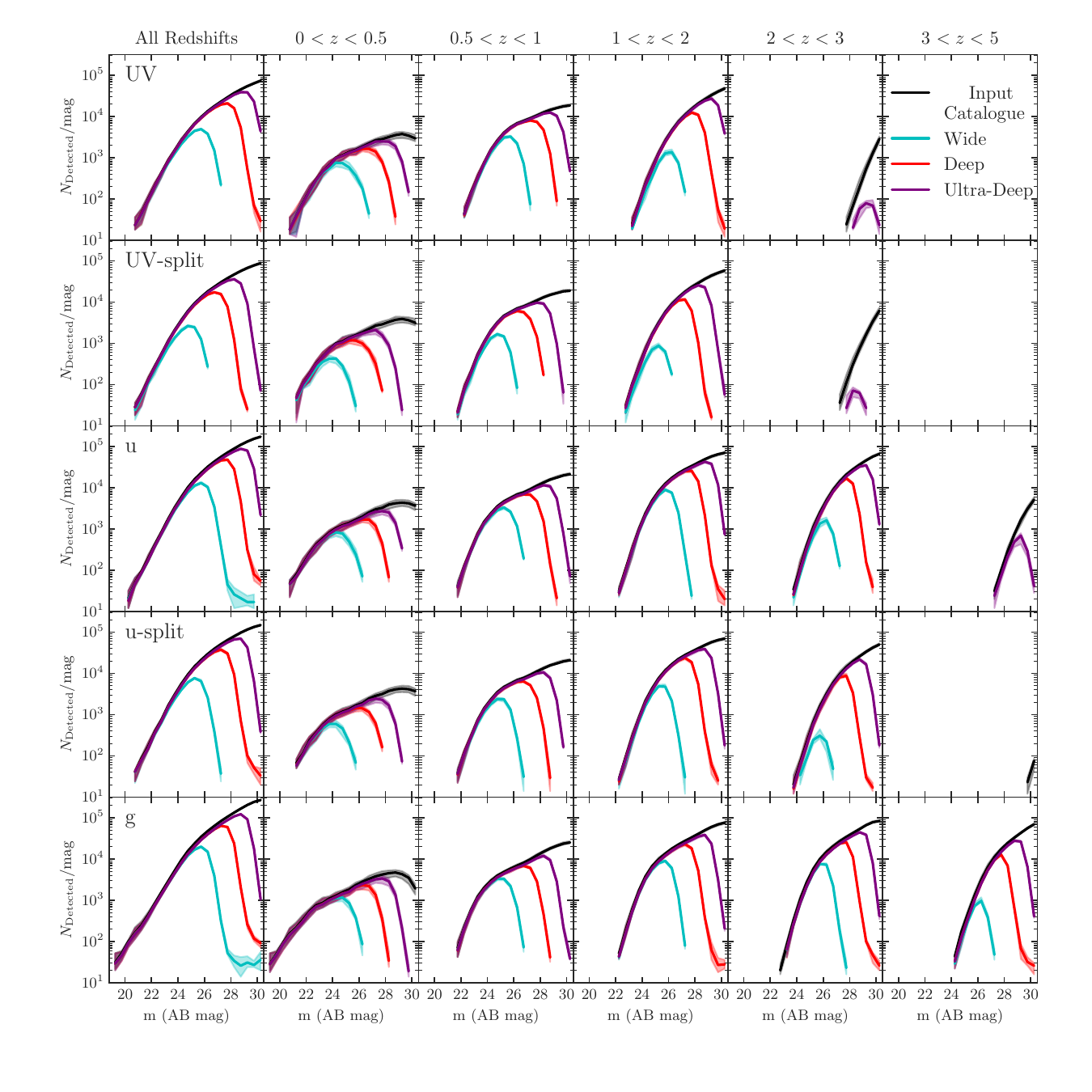}
\begin{center}
\caption{The number of galaxies that are detected by Source Extractor in one mock CASTOR pointing as a function of magnitude, for exposure times corresponding to the planned CASTOR Wide, Deep and Ultra-Deep surveys. The magnitude distribution of the input, simulated catalogue is also shown in black. 
The range shows the minimum and maximum fractions as measured in 6 separate pointings, with the solid line showing the median value.
Each row shows a different filter, and each column shows a specific redshift range (see titles).}
\label{fig:Ndetected_mag}
\end{center}
\end{figure*}

\begin{figure*}
\begin{center}
\hspace{-1cm}
\vspace{-1cm}
\includegraphics[scale=0.8]{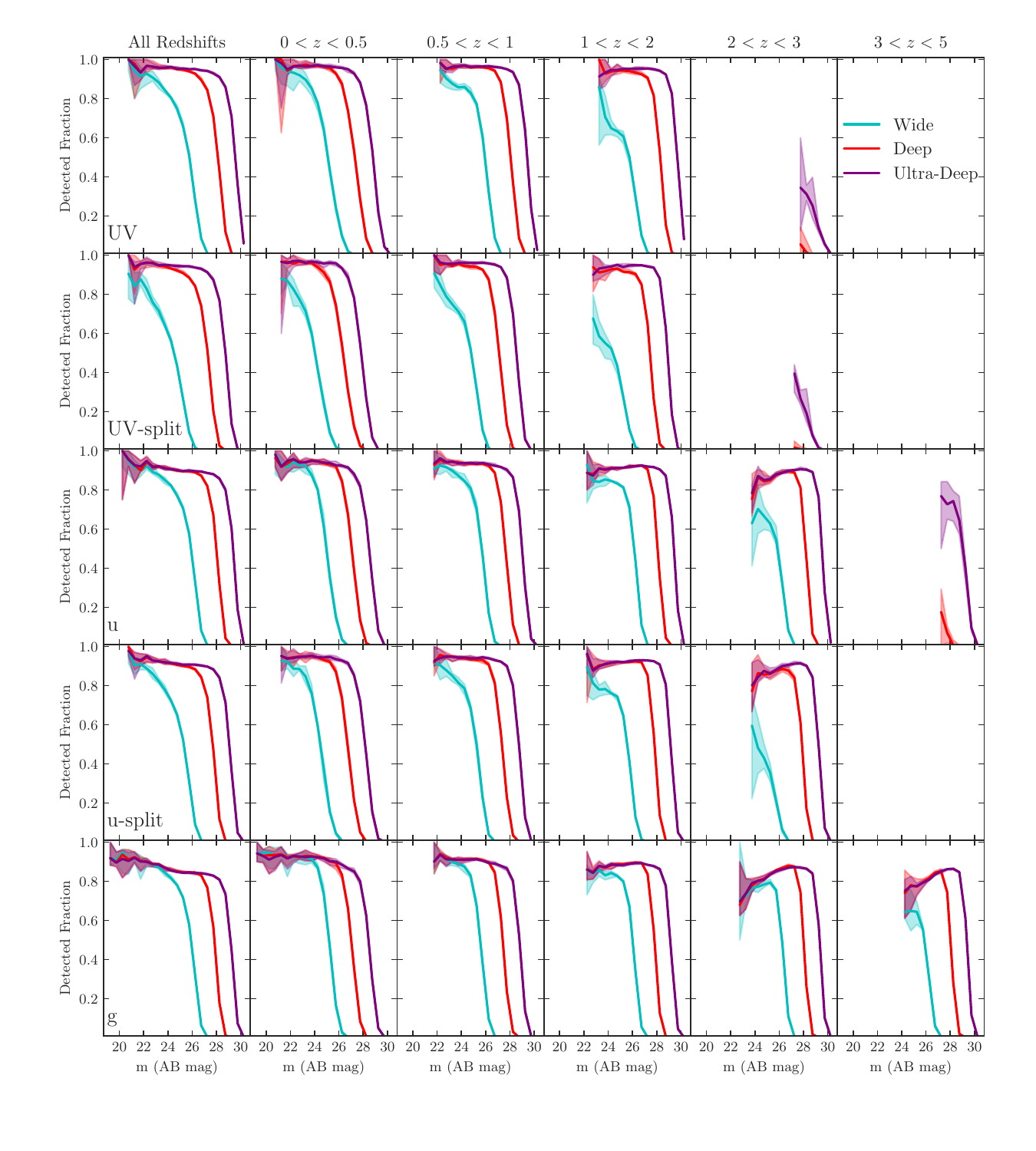}
\caption{The fraction of galaxies that are detected by Source Extractor relative to the input simulated catalogue as a function of magnitude, for exposure times corresponding to the planned CASTOR Wide, Deep and Ultra-Deep surveys. The range shows the minimum and maximum fractions as measured in 6 separate pointings, with the solid line showing the median value.
Each row shows a different filter, and each column shows a specific redshift range (see titles).}
\label{fig:FracDetected_mag}
\end{center}
\end{figure*}

To investigate the success of CASTOR in detecting extended galaxies, we consider mock images of six CASTOR pointings, taken from independent, non-overlapping regions of the input 2 deg$^2$ catalogue. This allows us to account for some cosmic variance. We extract and then match all detected galaxies within each of these six images, for each filter and survey exposure time.

In Figure \ref{fig:Ndetected_mag} we show the number of galaxies seen in one CASTOR FOV, as a function of magnitude, for each of the filters. We plot the number of galaxies detected for three exposure times, corresponding to the three CASTOR surveys: Wide (1000s, with 2000s for g), Deep (18,000s, with 36,000s for g), and Ultra-Deep (180,000s, with 360,000s for g). We compare these number counts to the input catalogue, showing the fraction of detected galaxies in Figure \ref{fig:FracDetected_mag}.
The 75\% completeness limit, measured across all redshifts from $0<z<5$, is given in Table \ref{tab:CompletenessLims}, alongside the faintest galaxy detected in each survey in each filter.

The Ultra-Deep survey with the largest exposure times clearly detects the most galaxies, with the largest detection fractions out to the faintest magnitudes. As shown in Table \ref{tab:CompletenessLims}, the 75\% completeness limits for this Ultra-Deep survey are 28.3--29.2 mag, 
with galaxies as faint as $\sim31$ mag detected. The filters are well matched in depth within this Ultra-Deep survey. 
The Deep survey will reach around 1.3--1.6 
magnitudes brighter than the Ultra-Deep survey, with a factor of 10 less exposure time. 

The significantly shallower Wide survey will be  75\% complete between 22.8--25.0 mag,
approximately 4--5
mag fainter than the Ultra-Deep survey, and will reach depths of up to 27.5 mag.
In this survey the decreased sensitivity of the UV-split and u-split filters is clear, resulting in a decreased completeness limit of up to 2
mag compared to their un-split counterparts. These narrower filters will play a key role in understanding the properties of the sources detected in CASTOR, providing further data points for SED fitting, as well as being combined with the un-split filter images to create even deeper detection images, however in general the UV, u and g filters will be the optimal filters for source detection in the Wide survey.

To further compare the performance of each survey, we list the number of galaxies detected in each CASTOR FOV in Table \ref{tab:GalaxyCounts}, as well as an estimation of the total number of galaxies in the full surveys by extrapolating this value across the survey areas. 
In one pointing, the Ultra-Deep survey is expected to detect an order of magnitude more galaxies than the Wide survey, of order 100,000 galaxies compared to $\sim10,000$. However, with its large survey area of 2227 deg$^2$, this will correspond to $\sim10^8$ detected galaxies across the whole Wide survey. The much deeper Ultra-Deep survey's small volume of only 4 CASTOR pointings (1 deg$^2$), will detect $\sim10^6$ galaxies, albeit down to much fainter magnitudes.
The intermediate Deep survey, covering 83 deg$^2$, will detect $\sim10^7$ galaxies.

In Figures \ref{fig:Ndetected_mag} and \ref{fig:FracDetected_mag} we also show the number of galaxies and detection fractions split into various redshift ranges, as taken from their input redshift from the catalogue.
The fraction of galaxies detected in the UV and UV-split bands at $z>2$ is very low. This is because the Lyman-limit at 912\AA\  is redshifted beyond the UV and UV-split filters, which extend to 3000\AA, at $z\geq2.3$. By $z\geq3.4$ this is redshifted beyond the u and u-split filters, which extend to 4000\AA. Thus similarly, we predict very few galaxies to be detected at $z>3$ within these filters. The g-band, covering up to 5500\AA, could detect galaxies out to $z\simeq5$, albeit with low detection rates. We investigate this further in Section \ref{sec:properties}.

\subsection{Completeness relative to galaxy properties}
\label{sec:properties}

\begin{figure*}
\begin{center}
\hspace{-1cm}
\vspace{-1cm}
\includegraphics[scale=0.8]{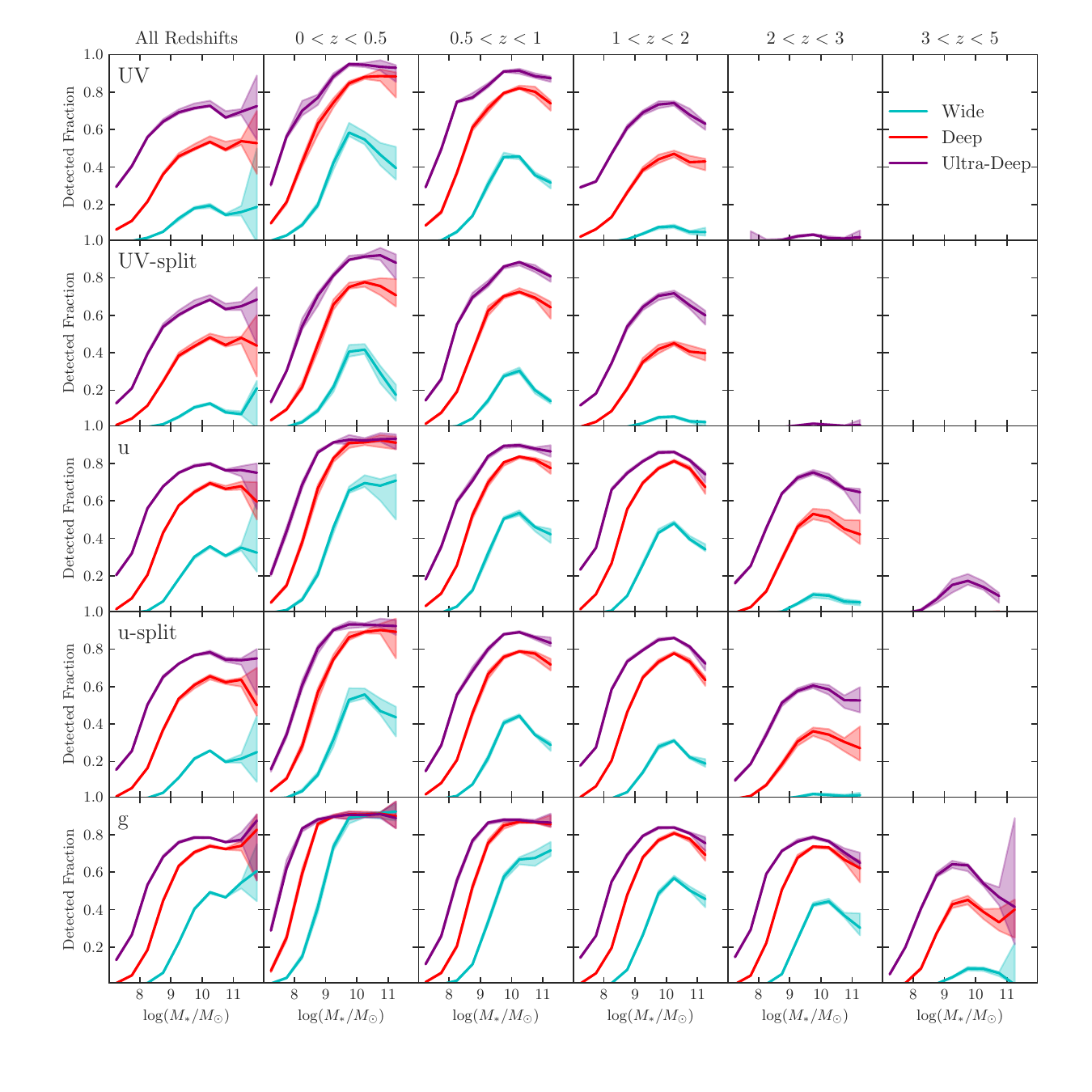}
\caption{The fraction of galaxies that are detected by Source Extractor relative to the input simulated catalogue, as a function of stellar mass, for each of the CASTOR surveys. The range shows the minimum and maximum fractions as measured in 6 separate pointings, with the solid line showing the median value.
Each row shows a different filter, and each column shows a specific redshift range (see titles).}
\label{fig:FracDetected_mass}
\end{center}
\end{figure*}

\begin{figure*}
\begin{center}
\hspace{-1cm}
\vspace{-1cm}
\includegraphics[scale=0.8]{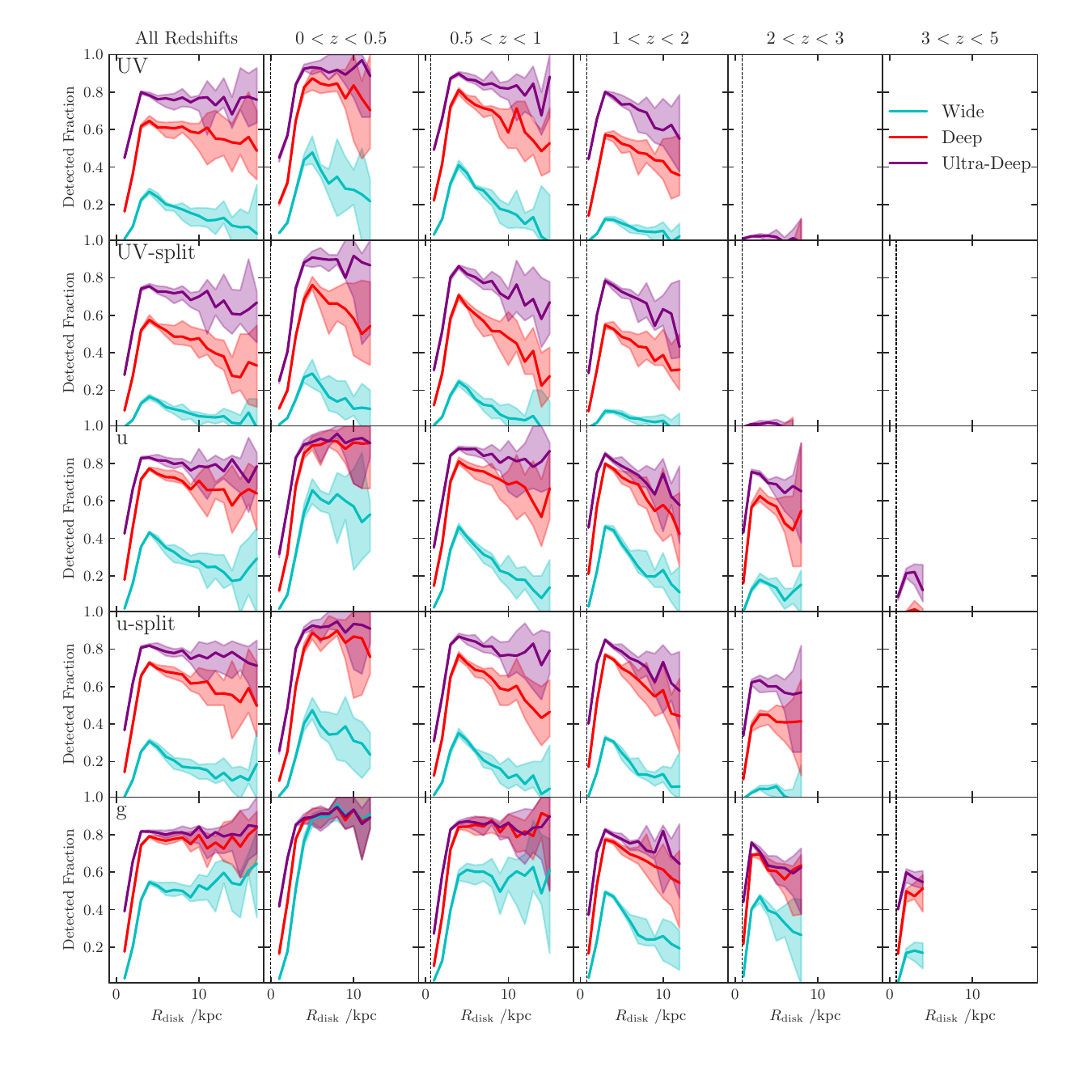}
\caption{The fraction of galaxies that are detected by Source Extractor relative to the input simulated catalogue, as a function of galaxy disk radius, for each of the CASTOR surveys. The range shows the minimum and maximum fractions as measured in 6 separate pointings, with the solid line showing the median value.
Each row shows a different filter, and each column shows a specific redshift range (see titles). The dashed vertical line shows the radius corresponding to the pixel size of $0\farcs1$ at the central redshift in the bin.}
\label{fig:FracDetected_radius}
\end{center}
\end{figure*}

\begin{figure*}
\begin{center}
\includegraphics[scale=0.8]{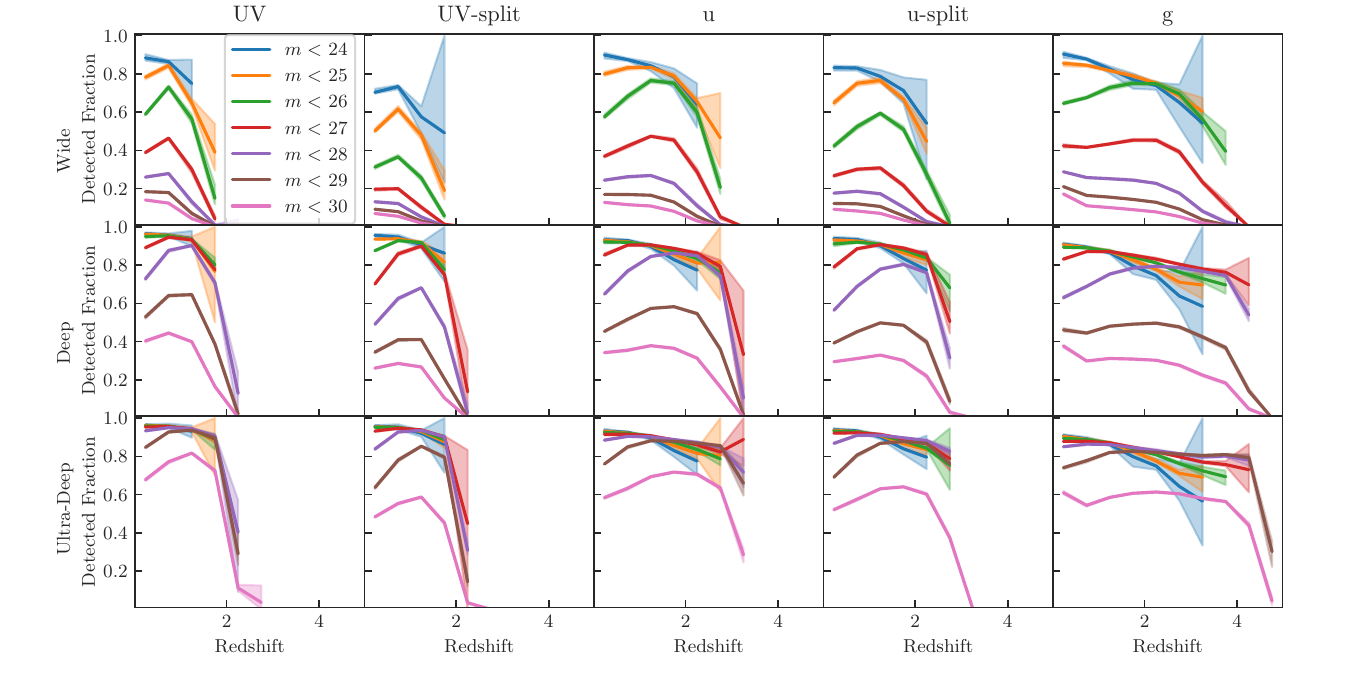}
\caption{The fraction of galaxies that are detected by Source Extractor relative to the input simulated catalogue,  as a function of redshift. Each of the curves shows a different limiting magnitude, from $m<24$ mag to $m<30$ mag, in the relevant filter.
The range shows the minimum and maximum fractions as measured in 6 separate pointings, with the solid line showing the mean value.
Each column shows a different filter. Each row shows a specific CASTOR survey.}
\label{fig:FracDetected_z}
\end{center}
\end{figure*}

We now investigate the detectability of galaxies in each CASTOR survey relative to the intrinsic galaxy properties from the input catalogue, for galaxies brighter than $\leq31$ mag. 
In Figure \ref{fig:FracDetected_mass} we show the fraction of input galaxies detected as a function of stellar mass for each survey. This is shown for each of the filters, and split into redshift bins. 
The detection fraction generally increases with galaxy stellar mass, with a plateau at large mass depending on the filter and exposure time. Generally, we expect the Wide survey to detect very few galaxies with $M_\ast\lesssim10^9M_\odot$. The Deep survey may detect a small fraction of galaxies down to $M_\ast\simeq10^8M_\odot$, while the Ultra-Deep survey may detect  
galaxies with $M_\ast\simeq10^8M_\odot$ at rates of $\sim50\%$, as well as a small fraction of galaxies down to $M_\ast\simeq10^7M_\odot$.

In Figure \ref{fig:FracDetected_radius} we show the equivalent figure of the fraction of galaxies detected as a function of galaxy disk radius. Galaxies with very small disk sizes of $\lesssim3$kpc have much lower rates of successful detection. 
The detection fractions also trend downwards for larger galaxies, particularly in the Wide survey, which is likely due to their lower surface brightnesses making them more difficult to observe.
The average detection fraction of disks with $\gtrsim3$kpc depends strongly on exposure time and filter, with the Wide survey having detection fractions of $\sim10\%$, up to $\sim80\%$ with the Ultra-Deep survey.

In Figure \ref{fig:FracDetected_z} we show the galaxy completeness as a function of redshift, for a range of various limiting magnitudes.
Clearly, fainter galaxy samples have lower detection fractions, except for in the Ultra-Deep survey which detects even $m<28$ mag galaxies with rates of $\sim100\%$
out to the highest observable redshifts.
In the Wide survey the detection fraction depends strongly on the magnitude of the sample, as does the trend with redshift. The brightest galaxies with $m<24$ mag are detected with fractions of $\gtrsim80\%$
in the UV, u, and g filters, out to $z\simeq3$ in the g band. This increases to $\sim100\%$
in the Deep and Ultra-Deep surveys. The limiting redshift in this case is due to the rarity of bright sources, which can be seen in Figure \ref{fig:Ndetected_mag}.

In the Deep and Ultra-Deep surveys, for a larger sample of galaxies down to much fainter magnitudes, $m<30$ mag, the detection fraction begins to decrease rapidly at the highest redshifts: at $z\gtrsim1.5$--2 in UV and UV-split, $z\gtrsim2.5$--3 in u and u-split, and at $z\gtrsim3.5$ in g. These decreasing detection fractions are caused by the Lyman-limit at 912\AA\ beginning to be redshifted into the filters, with galaxies at higher redshifts having minimal flux in these bands. The limiting redshift is also determined by the Lyman-limit, which is redshifted out of the UV filter at $z\geq2.3$, u at $z\geq3.4$, and g at $z\simeq5$.

\subsection{Magnitude recovery}
\begin{figure*}
\begin{center}
\vspace{-0.5cm}
\includegraphics[scale=0.8]{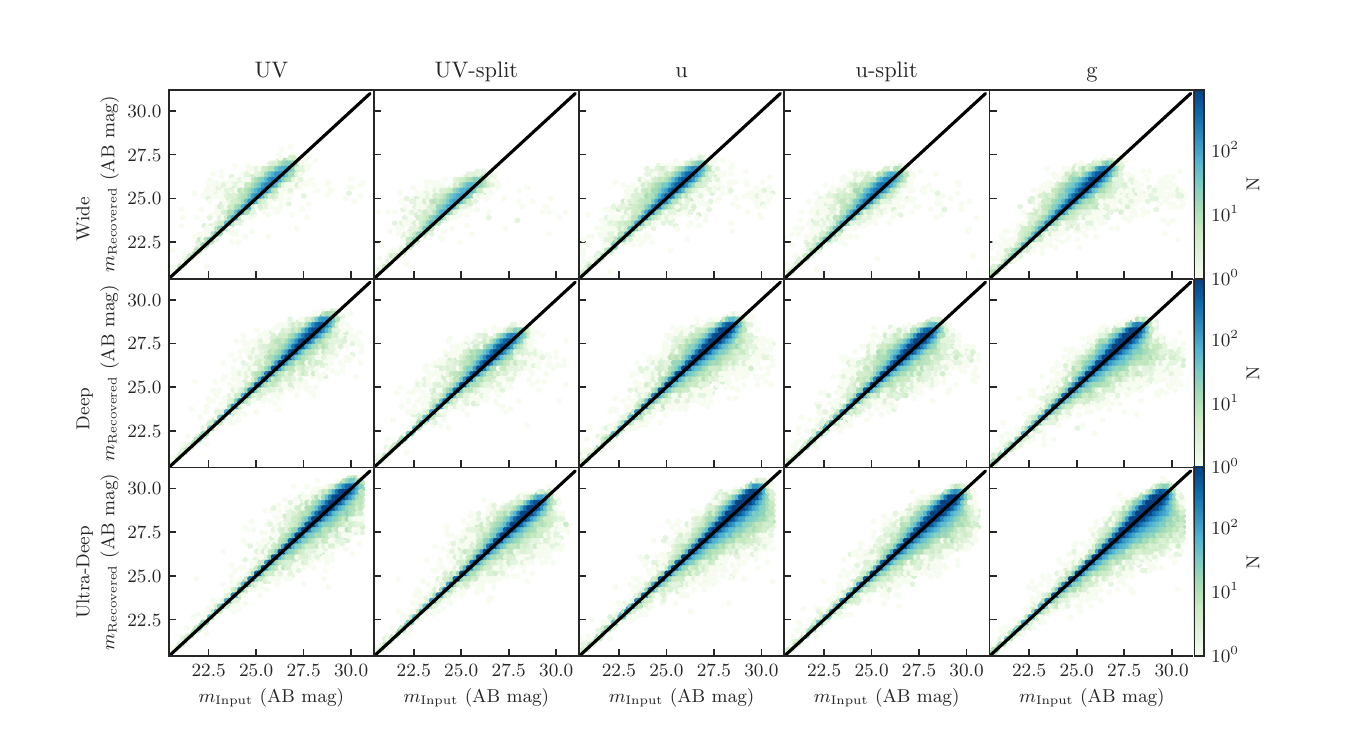}
\caption{The Source Extractor extracted magnitude for each of the detected galaxies, compared to the input galaxy magnitude from the theoretical input catalogue, in a given CASTOR FOV. Each column shows a different filter. Each row shows a specific CASTOR survey.}
\label{fig:InputOutput}
\end{center}
\end{figure*}

Finally we consider how the input and recovered magnitudes compare.
In Figure \ref{fig:InputOutput} we show the extracted ISO magnitudes from Source Extractor, compared with the input magnitude. We note that throughout this paper we have otherwise considered only the input theoretical magnitude.

We find that in the Wide survey, 95\% of all galaxies have an extracted magnitude within 
3.3\%, 4.1\%, 2.9\%, 3.2\% and 2.9\% 
of the input magnitude, for the UV, UV-split, u, u-split, and g filters respectively. 
These extracted magnitudes give accurate measurements of the input magnitudes, with a median bias of 0.6\% above the input magnitudes.
These are slightly improved in the Deep and Ultra-Deep surveys, to 
2.0\%, 2.2\%, 1.9\%, 2.0\% and 2.0\% 
for Deep,
and 
1.7\%, 1.8\%, 1.6\%, 1.7\% and 1.7\% 
for Ultra-Deep. 
The extracted magnitudes from both the Deep and Ultra-Deep surveys have a median bias of 0.4\% above the input magnitudes.
The low number of galaxies with larger discrepancies between the input and output magnitudes are very faint.

The extracted magnitudes are both accurate and precise, with median bias $\lesssim0.6\%$, and with 95\% of galaxies having magnitudes correct to within $\lesssim4\%$, for all filters in the three surveys. 
Overall, the CASTOR surveys can accurately measure galaxy magnitudes.

\section{Discussion}
\label{sec:Discussion}
\subsection{Synergy with Euclid and Roman}

CASTOR's wide-field survey capabilities and UV coverage will play an indispensable role in the new era of wide-field near-infrared extragalactic surveys, alongside Roman \citep{Spergel2013, Spergel2015} and Euclid \citep{Laureijs2011,Racca2016}. These space-based observatories have fields of view of the order of hundred times bigger than Hubble's (0.54 deg$^2$ for Euclid and 0.28 deg$^2$ for Roman) and possess angular resolutions comparable to that of Hubble (with a mirror diameter of 1.2 m for Euclid and 2.4 m for Roman), similar to those of CASTOR (0.25 deg$^2$ and $D=1$ m).
Roman and Euclid focus on the near-infrared and optical, covering wavelengths up to $\sim2.2\mu$m. However,  as these will provide no wavelength coverage below $\sim5000$\AA, CASTOR will play a key role with its wavelength coverage that extends beyond the optical to the near UV.
Together, these three observatories are set to transform galaxy and cosmological surveys. 
Harnessing the synergy between CASTOR and other observatories will be the key to maximizing the scientific return across all observatories.

In particular, extragalactic surveys in the UV, u and g bands will enable diagnostics that require the additional UV coverage (e.g. UVJ color selection) and deliver important multi-wavelength diagnostics for star-forming galaxies \citep[e.g.][]{Shen2023,Mehta2023}, which will be highly desirable in identifying sources in the planned Roman Core Community Surveys and Euclid Deep Fields.
In addition, CASTOR grism surveys can also deliver fast emission-line diagnostics (e.g. BPT diagram, \citealt{Baldwin1981}), which can reliably and efficiently classify galaxies and AGN in the era of wide-field surveys.

The CASTOR Wide survey is planned to overlap the footprint of the Roman HLS, which will also be mapped by Euclid and Rubin. The CASTOR Deep and Ultra-Deep fields are also planned to be conducted in fields with deep Roman, Euclid, and/or Rubin coverage, to optimise the science output of these surveys.
Roman is expected to reach depths of 26--27 mag across the HLS, with a potential for an embedded field going $\sim 1.5$ mag deeper \citep{Spergel2015}.
Euclid's Wide survey will reach depths of 24.5--26 mag, with deeper individual fields reaching $\sim2$ mag deeper \citep{Scaramella2022}.
The 10-year LSST survey is predicted to reach 25--27.5 mag in depth \citep{Ivezic2019}. The equivalent point source depths for the CASTOR Wide survey from Table \ref{tab:CompletenessLims} are 26--28 mag, showing that CASTOR is well-matched in depth to these other wide-field surveys. 
The CASTOR Deep survey will reach 28--29 mag, similar to the deeper Roman and Euclid fields, while the Ultra-Deep field will exceed these by extending to $\sim30$ mag (Table \ref{tab:CompletenessLims}).
Together these surveys will provide a comprehensive, high-resolution coverage of the sky across the UV to the near-infrared.

\section{Conclusions}
\label{sec:Conclusions}
The Cosmological Advanced Survey Telescope for Optical and UV Research (CASTOR) is a planned Canadian-led space telescope designed to study the Universe in the UV and optical. 
A core component of the planning for CASTOR is the preparation of a suite of mission planning and science simulation tools, the Finding Optical Requirements and Exposure times for CASTOR (FORECASTOR) project.
In this work we present the publicly available CASTOR image simulator\footnote{\url{https://github.com/CASTOR-telescope/ImageSimulator}}, a generator which simulates extragalactic imaging for CASTOR observation planning.
This generator is built upon the framework of the GalSim Python package, and creates images containing sources from simulated light-cones from the Santa Cruz Semi-Analytic Model.

Three main extragalactic surveys planned for the mission are the Wide, Deep, and Ultra-Deep Legacy Surveys, spanning 2227, 83 and 1 deg$^2$, respectively. These surveys will reach total exposure times of 1000, 18,000 and 180,000s respectively in the UV, UV-split, u and u-split filters, with twice the exposure time in the g filter. In this paper we present mock images of each of these surveys, and investigate the detectability of galaxies at these planned exposure times.
These images are publicly available\footnote{\url{https://doi.org/10.11570/24.0003}}, to enable further investigations of the extragalactic science achievable with CASTOR.
The $\sim0\farcs15$ resolution of CASTOR will result in crisp UV images, which will significantly out-perform GALEX with its $\sim5''$ that resulted in significant galaxy blending.
This is also a higher resolution than for Rubin, which is limited to $\sim0\farcs8$ due to atmospheric seeing.

We extract sources from the images and compare those to the input source catalogue, in order to estimate the completeness of each CASTOR survey.
For point sources, we predict that the Wide survey will be 75\% complete down to 26.2--27.2 mag, with the Deep survey extending to 28.2--28.8 mag, and the Ultra-Deep survey reaching to 29.5--30.2 mag. 
The five filters, UV, UV-split, u, u-split, and g, are generally well matched in depth for each survey.
For galaxies, we expect that the faintest detectable galaxy discovered in the Wide survey will have $\sim 27.5$ mag, 
with Deep reaching to $\sim29.5$ mag 
and Ultra-Deep discovering some faint sources out to $\sim31$ mag.
The recovered magnitudes are in good agreement with those from the input catalogue, with median bias $\lesssim0.6\%$, and with 95\% of galaxies having magnitudes correct to within $\lesssim4\%$, for all filters in the three surveys. 
For the completeness relative to stellar mass, we expect the Wide survey to detect very few galaxies with $M_\ast\lesssim10^9M_\odot$. The Deep survey may detect galaxies down to $M_\ast\simeq10^8M_\odot$, while the Ultra-Deep survey may detect galaxies down to $M_\ast\simeq10^7M_\odot$. Galaxies with disk radii of $\lesssim3$ kpc will be the most difficult to detect.

For a single 0.25 deg$^2$ pointing, the Wide survey is expected to detect of order $\sim10,000$ galaxies, with $\sim100,000$ galaxies seen in one pointing of the Deep and Ultra-Deep surveys. 
Extending these counts to the full survey areas, we expect the Wide survey to detect $\sim10^{8}$ galaxies, with $\sim10^{7}$ galaxies in the Deep survey and $\sim10^{6}$ in the Ultra-Deep survey. 

In the UV and UV-split filters, CASTOR will detect galaxies out to $z\lesssim2$. In the u and u-split filters, galaxies can be detected to $z\lesssim3.5$, while galaxies out to $z\lesssim5$ can be detected in g, due to the Lyman-limit.

Overall, CASTOR's resolution and large field of view will result in the detailed detection of hundreds of millions of galaxies in the UV and optical. The Wide survey will be a prime complement to the Euclid and Roman large optical/infrared surveys, providing critical UV coverage to equivalent depths that will be vital for SED fitting. It will also enrich the science capable with  Rubin, with CASTOR's higher resolution resolving blends and providing accurate galaxy shape measurements that will be critical to understanding lensing and galaxy evolution.


\section*{Acknowledgements}
We sincerely thank the referee for their supportive and helpful feedback on this project.

MAM acknowledges support by the Laboratory Directed Research and Development program of Los Alamos National Laboratory under project number 20240752PRD1.

This research used the Canadian Advanced Network For Astronomy Research (CANFAR) operated in partnership by the Canadian Astronomy Data Centre and The Digital Research Alliance of Canada with support from the National Research Council of Canada, the Canadian Space Agency, CANARIE, and the Canadian Foundation for Innovation.

This paper made use of Source Extractor \citep{SExtractor}, and Python packages and software
AstroPy \citep{Astropy2013,Astropy2018},
GalSim \citep{2015A&C....10..121R},
Matplotlib \citep{Matplotlib2007},
NumPy \citep{Numpy2011},
Pandas \citep{reback2020pandas}, 
Photutils \citep{photutils},
SciPy \citep{2020SciPy-NMeth}, and 
Seaborn \citep{Waskom2021}.

\section*{Data Availability}
The mock CASTOR survey images generated in this work are available on CANFAR, at \url{https://doi.org/10.11570/24.0003}. The CASTOR image simulator, including the currently estimated CASTOR filter bandpasses and PSFs, is available at \url{https://github.com/CASTOR-telescope/ImageSimulator}.
The simulated lightcones used in this work are not publicly available, but will be shared on reasonable request to the corresponding author.
\bibliographystyle{mnras}
\bibliography{manuscript.bib}{}

\appendix
\section{The effect of noise assumptions on our completeness limits}
\label{sec:noiseAppendix}
In our main set of images, we assumed a read noise of 3.0 e-/pixel, that is applied once per 3000s block of exposure time, a dark current of 0.002 e-/pixel/s, and a `Medium' Zodiacal light background. These assumptions are based on our current estimates of the CASTOR detector performance, however these may vary, and will also depend on the decided readout strategies, as well as the position on the sky. 

\begin{figure*}
\begin{center}
\vspace{-1cm}
\includegraphics[scale=0.8]{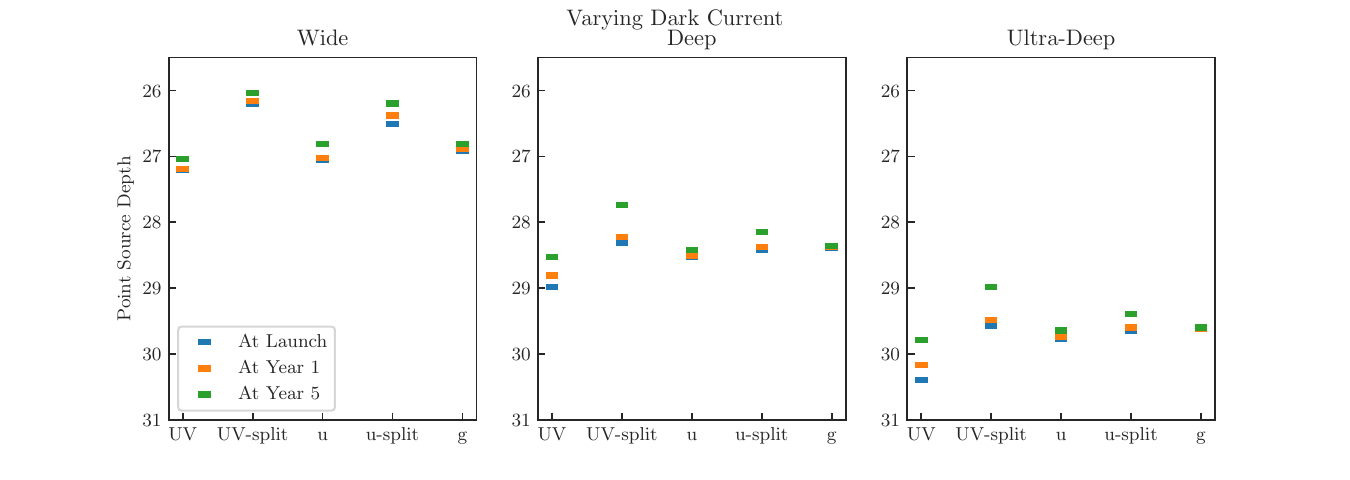}
\includegraphics[scale=0.8]{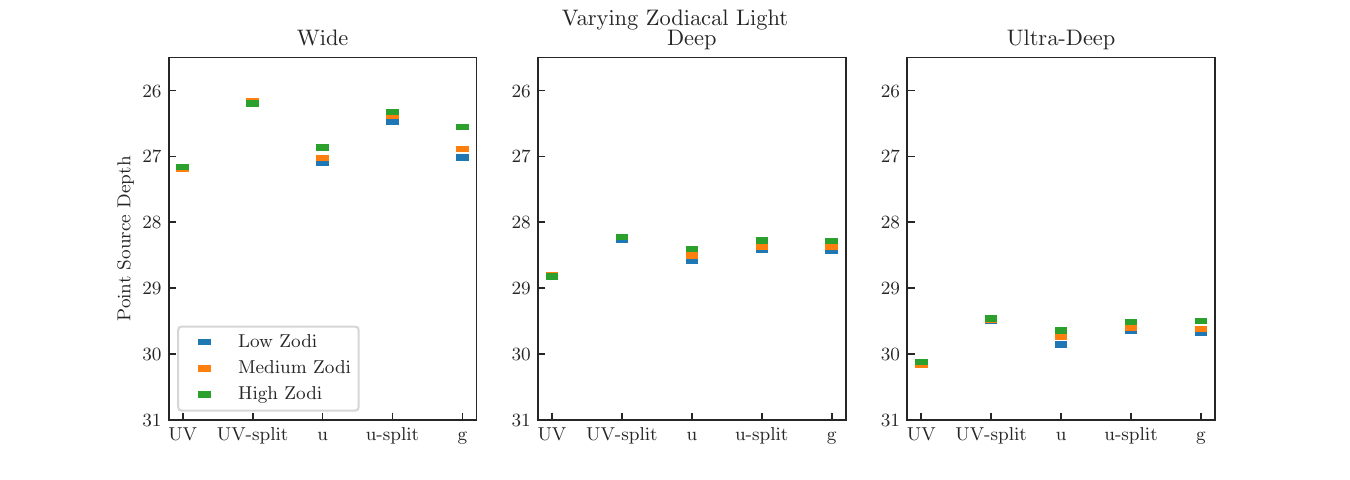}
\caption{The 75\% completeness limits for point sources in each of the five filters, for each of the Wide, Deep and Ultra-Deep surveys (left to right columns). Top panels: Variation due to a changing dark current over time, from 0.0001 e-/pixel/s at launch, 0.002 e-/pixel/s after 1 year, to 0.01 e-/pixel/s after 5 years. Bottom panels: Variation with a different assumed zodiacal light level, in Johnson V: 23.3 mag for `Low', 22.7 mag for `Medium', and 21.1 mag for `High'. 
}
\label{fig:NoiseVariation}
\end{center}
\end{figure*}

The dark current of the CMOS detector is expected to increase linearly in time, as for CCDs \citep{ACShandbook}. The initial dark current is expected to be 0.0001 e-/pixel/s, increasing to 0.01 e-/pixel/s at the end of five years.
As the three main surveys are expected to be carried out across the first two years of the mission, we assume an average of 0.002 e-/pixel/s in our images. In Figure \ref{fig:NoiseVariation} we show how the 75\% completeness limit for point sources changes with time due to the varying dark current, from 0.0001 e-/pixel/s at launch to 0.01 e-/pixel/s at the end of five years.
The UV and UV-split filters will have the largest degradation in sensitivity over time, with up to a 0.6 mag decrease in the point source completeness limit for a 180,000s (Ultra-Deep survey-depth) image after 5 years.

Throughout our work we assume the `Medium' zodiacal light estimate as used in the HST ETC. 
In the HST ETC the template zodiacal spectrum is normalized to three Johnson V magnitudes:
23.3 mag for `Low', 22.7 mag for `Medium', and 21.1 mag for `High'. In Figure \ref{fig:NoiseVariation} we show how the 75\% completeness limit for point sources changes under these various Zodiacal assumptions (using the standard 0.002 e-/pixel/s dark current).
The largest effect is seen for the g filter in the Wide survey, with depth varying by 0.5 mag between the Low and High cases.
Varying from Low to High zodiacal light decreases the completeness limit by 0.2 mag in the u filter in all surveys, and 0.15 mag in the u-split filter, while the UV filters are not significantly affected.

\bsp
\label{lastpage}
\end{document}